\documentclass{article}
\usepackage{amsmath, amsfonts, amssymb, siunitx, graphicx, setspace, subcaption, booktabs}
\usepackage[separate-uncertainty=true]{siunitx}
\usepackage[margin=0.75in]{geometry}
\usepackage[hyphens]{url}
\usepackage[breaklinks=true]{hyperref}

\newcommand{\diff}{\text{d}}

\DeclareSIUnit{\Wh}{Wh}
\DeclareSIUnit{\month}{mo.}
\DeclareSIUnit{\day}{day}
 \DeclareSIUnit\bar{bar}
\title{Techno-economic optimization of hybrid steam-electric energy systems with excess heat utilization and reserve market participation}

\author{
  Olav Galteland, Jacob Hadler-Jacobsen, and Hanne Kauko \\
  \normalsize Department of Thermal Energy, SINTEF Energy Research, Norway
}

\begin{document}
\maketitle

\begin{abstract}
This study investigates the economic viability and optimal configuration of a hybrid industrial energy system combining an electrode boiler, steam accumulator, and battery energy storage system (BESS). This study optimizes system operation for a specific configuration to minimize net energy costs, defined as energy costs minus profits from price arbitrage and reserve markets. The optimization uses load shifting, peak shaving, and frequency containment reserve market participation with hourly 2024 data from Norway and Germany. Net present value (NPV) analysis was performed to determine the most cost-efficient energy storage configurations. The results show that current investment costs favor steam accumulators over BESS in both countries. However, a reduction in BESS cost will make batteries economically competitive, particularly in Germany, where high price volatility and power-based grid tariffs provide stronger incentives for load shifting and peak shaving. Participation in the FCR market accounted for a 17\% and 7\% reduction of the net energy costs in Norway and Germany, respectively. Utilization of excess heat, through inlet water preheating, further reduced the net energy costs. Sensitivity analyses confirm that investment costs, especially for BESS, strongly influence optimal system design. These findings offer guidance for industrial flexibility investments across diverse electricity markets.
\end{abstract}

\section{Introduction}

The electricity grids are under stress in many parts of the world \cite{IEA2023Grids}. This trend is driven by rising electricity demand resulting from the electrification of transport and industry and the emergence of new power-intensive sectors. It is also fueled by the growing integration of intermittent renewable sources, primarily wind and solar \cite{IEA2024WEO}. To avoid faults in the electricity grid, either massive economic investments into more grid capacity must be made to meet the increasing peak demands, or the electricity grid must be utilized in a smarter way, for example by utilizing flexibility through energy storage. The electricity markets are designed to incentivize asset owners to offer flexibility to the grid operators through price arbitrage, reserve market participation, and peak tariffs. Price arbitrage in the electricity grid is the act of buying electricity during low-cost periods, and selling during high-cost periods. This provides the asset owners with profits, while the grid operators gest help maintaining a stable grid. The reserve market is another tool, with several products with varying response times and durations. In this work, we will focus on the frequency containment reserves (FCR) product, which has a response time of 30 seconds and a duration of 15 minutes \cite{entso-e}. 

Electricity is however just one part of the energy puzzle. Approximately half of the European energy demand is in the form of heating, with space heating and industrial heating as the most dominant heating demands \cite{heatroadmapeurope2017}. Approximately 20\% of the energy consumption in the EU was used for industrial heating \cite{EERA202XIndustrialTES}, in processes that often generate large amounts of excess heat. Approximately 10\% of the primary energy consumption in Norway is dissipated as excess heat, much of it above \SI{100}{\celsius} \cite{HighEFF2023}. Excess heat from industry has a high potential to be utilized in district heating networks to support the decarbonization of space heating and industrial heat supply \cite{manz2021decarbonizing}, and to reduce the impact of decarbonization on the grid. 

There is additionally a huge flexibility potential in the coupling between renewable-based power grid and thermal energy systems. Integrated and efficient energy systems, with interaction between the different energy carriers and end-user sectors, are highlighted as key strategies for developing a competitive and sustainable energy supply for Europe~\cite{EUintegration2020}. Considering the high energy demand for heating, a flexible interaction between the electricity grid and the heating sector is particularly important for an affordable energy transition~\cite{mathiesen2009, bloess2018}.  Industrial steam generation is especially interesting, since (i) majority of its demand is covered by fossil fuels~\cite{whitepaperHTHP2020} and (ii) industrial steam generation units often consume large amounts of power, with high potential for providing flexibility services.

The economic viability of flexible hybrid steam and electricity networks depends critically on the capital and operational costs of electrode boilers, steam accumulators, and battery energy storage systems (BESS). Cost reductions for electrode boilers and steam accumulators are primarily driven by material costs (e.g., steel). In contrast, BESS have experienced a dramatic decrease in costs over the past decade, driven by rapid technological advancements and economics of scale. Approximately 60\% of the BESS cost is for battery cells, while the remaining 40\% is attributed to e.g. converters, electronics, cooling system, and container. The lithium-ion cell cost reduction closely follows Wright's Law, which posits an inverse power-law relationship between cost and cumulative production volume. The electric vehicle (EV) market has been a primary driver for this trend, with global EV battery deployment reaching \SI{750}{\giga\Wh} in 2023 \cite{IEA2024Batteries}. The average lithium-ion battery cell prices plummeted from approximately \SI{550}{USD\per\kilo\Wh} in 2010 to below \SI{60}{USD\per\kilo\Wh} in 2025 \cite{IEA2024Batteries, SPGlobalMobility2024BatteryPrices, BNEF2024BatteryPrices}. The International Energy Agency (IEA) projects continued cost declines for BESS in the coming years, driven by further innovation and increasing market volume, a trend not anticipated for the more mature electrode boiler and steam accumulator technologies \cite{IEA2024Batteries}. As the cost of BESS continues to decrease, it is expected that large amounts of these will be used for ancillary services. The demand for ancillary services is also expected to increase with increasing amount of renewable energy sources. These effects are expected to influence the future profits from ancillary services, but their impact on the profitability of participating in reserve markets remains unclear \cite{seifert2024coordinated, kempitiya2020AIbidding}. 

Previous research has explored the benefits of combining thermal energy storage with BESS in industrial settings, demonstrating its potential for cost reduction and grid stabilization \cite{berg2024industrial, foslie2023integrated}. However, few studies have explicitly compared the economic viability of combining a steam accumulator and BESS in the context of both FCR market participation and price arbitrage, particularly across diverse electricity market structures such as those of Norway and Germany. Norway's electricity system is dominated by hydropower, leading to relatively low and stable electricity prices and significant capacity for providing frequency regulation services, but with potential for grid congestion in certain areas. Germany, on the other hand, has a higher proportion of intermittent renewable energy sources (wind and solar), resulting in greater price volatility in the spot market. This difference in price dynamics presents contrasting opportunities for price arbitrage and FCR participation, making a comparative analysis valuable for understanding the economic viability of BESS and steam accumulators in different market contexts. 

This study investigates energy flexibility in an industrial setting using a factory as a case study. The factory's annual steam demand is \SI{4.6e6}{\kilogram}. This amount of steam corresponds to \SI{3.5}{\giga\Wh} of thermal energy for steam produced from water heated from \SI{10}{\degree\celsius}. The steam demand timeseries used in this work has been provided by an industry site in central Norway. We analyze the integration of both a steam accumulator and a BESS to enable the factory to meet its steam requirements while also participating in FCR markets and exploiting electricity price arbitrage. The objective is to determine the optimal configuration of an electrode boiler, BESS, and steam accumulator for this industrial site. We use a two-stage optimization procedure, first minimizing net energy costs, defined as energy costs minus profits from FCR participation, for a given system configuration. From a set of configurations, we use net present value (NPV) analysis to find the most cost-efficient system configuration. Net energy costs are calculated using hourly FCR and spot price data from 2024 for Norway and Germany. The net energy costs for the year 2024 are then extrapolated to the system's assumed 15-year lifetime to estimate the system's NPV. This paper offers a unique contribution by comparing the performance of steam accumulators and BESS in an industrial setting, considering both technical and economic factors within a comprehensive NPV analysis, using data from two different markets. The choice of Norway and Germany as data sources reflects key differences in their electricity market structures and energy mixes, which significantly impact the potential for energy flexibility and the profitability of storage technologies. Furthermore, analysis of utilization of excess heat for preheating inlet water and anticipated BESS price trajectories provide valuable insights for policymakers, energy planners, and industrial stakeholders.

\section{System description}

\begin{figure}
    \centering
    \includegraphics[width=0.75\linewidth]{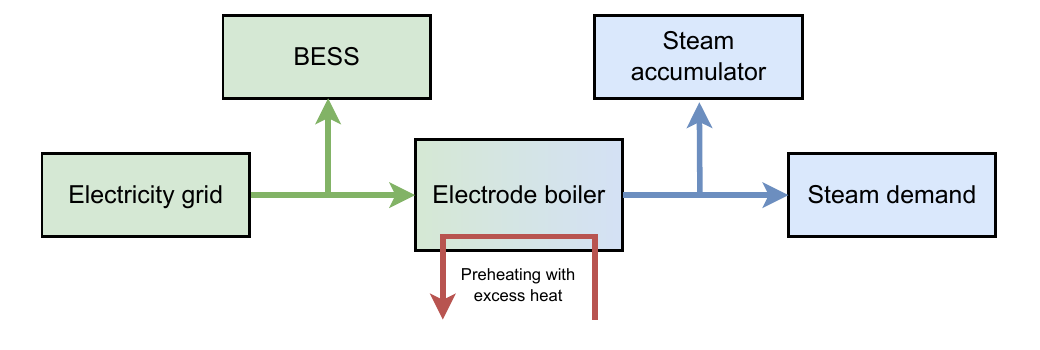}
    \caption{Schematic of the hybrid steam and electricity network. On the electricity side there is the electricity grid and BESS, and on the steam side there is a steam accumulator and steam demand. Connecting the two energy carriers is the electrode boiler, which can have inlet water preheated by excess heat.}
    \label{fig:network}
\end{figure}

In this work we will analyze a simplified industrial steam and electricity network. The network is connect to the electricity grid and a BESS on the electricity side. On the steam side, the network is connected to a steam accumulator and a factory with a steam demand. Between the two networks we have an electrode boiler, connecting the two energy carriers.  We will also consider preheating of the inlet water to the electrode boiler with excess heat. The network is illustrated in figure~\ref{fig:network}. 

\subsection{Electrode boiler and accumulator}
\label{sec:system:steam}
We will simplify the thermodynamic description of the electrode boiler and the steam accumulator to only be a function of their mass content of steam, and derive charge and discharge efficiencies and the self-discharge rate. The operating temperature and pressure will be assumed to be constant at $T$ and $p$. The electrode boiler and the steam accumulator will be considered as components for energy conversion and storage, meaning that the thermodynamics and fluid dynamics of two-phase steam and water will be simplified greatly. We will consider the steam to be saturated and dry. In this following section, we will derive the constraints of the electrode boiler and steam accumulator. The purpose is to have a convex description of the dynamics of the electrode boiler and the steam accumulator which can be utilized in energy system optimization.

The steam mass balance in the steam accumulator is
\begin{equation}
    \frac{\partial M_\text{sa}}{\partial t} = \dot{m} - \dot{m}_\text{loss}
\end{equation}
where the first term represents the mass flow rate of steam into the container from the steam network, and the second term represent the mass loss rate of steam due to heat loss from the container and the pipes to the surrounding environment. This is usable steam lost due to condensation.

The mass rate of steam production in the electrode boiler is
\begin{equation}
    \dot{m}_\text{prod} = \frac{P_\text{eb}}{\Delta h_\text{tot}},
\end{equation}
where $P_\text{eb}$ is the electric power input to the electrode boiler, and this steam is directly inserted to the steam network, as the electrode boiler is simplified to have no storage capacity. Here $\Delta h_\text{tot}$ is the total specific enthalpy difference from water at the inlet temperature and pressure ($T_0, p_0$) to the steam at operating temperature and pressure ($T, p$). This is the energy necessary to produce steam. The inlet temperature can be increased by preheating, and we will use this to investigate the utilization of excess heat. The total specific enthalpy is equal to
\begin{equation}
    \Delta h_\text{tot}(T,p) = h(T,p) - h(T_0, p_0) = \int_{T_0}^{T_b}c_p^w\diff T + \int_{T_b}^Tc_p^s\diff T+\Delta h_\text{vap}
\end{equation}
which is the sum of sensible heating of water and steam, and latent heat of vaporization. The symbols $c_p^w$ and $c_p^s$ are the specific heat capacities for water and steam respectively. The operating temperature is $T$, boiling temperature is $T_b$ and the inlet temperature is $T_0$.

The steam mass loss rate is caused by three terms, heat loss in the piping to and from the steam accumulator (charge and discharge, respectively), and heat loss in the steam accumulator,
\begin{equation}
    \dot{m}_\text{loss} = \frac{1}{\Delta h_\text{tot}} \left(q_\text{loss}^+ + q_\text{loss}^- + q_\text{loss}^\text{tank}\right).
\end{equation}
The heat losses in the pipe and steam accumulator are modeled as
\begin{equation}
    q_\text{loss}^+ = 2\pi\frac{\lambda^\text{pipe}}{\delta^\text{pipe}}L^+r (T-T_a)\quad\text{and}\quad q_\text{loss}^\text{tank} = 2\pi\frac{\lambda^\text{tank}}{\delta^\text{tank}}(HR+R^2) (T-T_a)
\end{equation}
where $L^+$ is the length of the steam pipe for charging, and $r$ is the radius of the steam pipe. An equivalent expression is used for the heat loss during discharge, with a pipe length of $L^-$. $H$ and $R$ are the length and radius of the steam accumulator, which is simplified to be cylindrical. The thickness and thermal conductivity of the insulation are ($\delta^\text{pipe}$, $\lambda^\text{pipe}$)  and ($\delta^\text{tank}$, $\lambda^\text{tank}$) for the pipe and tanks, respectively.

The steam mass capacity of the steam accumulator is
\begin{equation}
    M_\text{sa}^{\max} = \rho V_{sa} = \pi\rho H R^2
\end{equation}
where $\rho$ is the mass density of the steam and $V_{sa}$ is the volume of the steam accumulator. The mass loss due to heat loss in the pipe is analogous to the concept of charge and discharge efficiencies in BESS, and the mass loss due to heat loss in the tanks is analogous to the concept of self-discharge. The mass balance in the steam accumulator can be rewritten on this form,
\begin{equation}
     M_\text{sa}^{t+\Delta t} = \left(\eta^+_\text{sa}\dot{m}^{+,t} - \frac{\dot{m}^{-,t}}{\eta^-_\text{sa}}\right)\Delta t + (1- \varepsilon_\text{sa}\Delta t) M_\text{sa}^t
\end{equation}
where $\eta^+_\text{sa}$ and $\eta^-_\text{sa}$ are the charge and discharge efficiencies, respectively, and $\varepsilon_\text{sa}$ is the self-discharge rate. The mass flow rate, $\dot{m}=\dot{m}^+-\dot{m}^-$, is split into two terms, one for charging and one for discharging, where $\dot{m}^->0$ and $\dot{m}^+>0$. The production of steam will be assumed to be completely efficient, meaning that the electrical energy input to the electrode boiler is completely converted to thermal energy. We define the charge and discharge efficiencies as the ratio of mass flow rate at the outlet and inlet of the piping. This can be reformulated to the ratio of the heat flow rate at the outlet and inlet,
\begin{equation}
    \eta^+_\text{sa} = \frac{Q_\text{in}^+-q_\text{loss}^+}{Q_\text{in}^+}=\left(1-\frac{q_\text{loss}^+}{\dot{m}^+\Delta h_\text{tot}}\right),
    \label{eq:sa_round-trip_efficiency}
\end{equation}
where we have used that the heat at the outlet is equal to the heat at the inlet minus the losses. An equivalent expression is found for the discharge efficiency,
\begin{equation}
\eta^-_\text{sa} = \frac{Q_\text{out}^-}{Q_\text{out}^--q_\text{loss}^-} = \left(1 - \frac{q_\text{loss}^-}{\dot{m}^-\Delta h_\text{tot}}\right)^{-1}
\end{equation}
Similarly, we define the self-discharge rate as the mass loss rate due to heat loss in the tank divided by the mass of steam in the tank. The mass loss rate due to heat loss in the tank is equal to the heat loss divided by the total enthalpy,
\begin{equation}
    \varepsilon_\text{sa} = \frac{q_\text{loss}^\text{tank}}{M\Delta h_\text{tot}}.
    \label{eq:sa_self-discharge_rate}
\end{equation}
We will use these relations to compare the performance of steam accumulators and batteries as energy storage devices. 

The charge and discharge rates (C-rate) of an electrode boiler combined with a steam accumulator, analogously to C-rate of batteries, depends on power rating of the boiler, the pressure drop in the steam network, and the size of the steam piping. However, in this work we will assume the power rating of the boiler to be the limiting factor, meaning that in the system optimization we will not constrain the mass flow rate in the steam pipes.

\subsection{Battery energy storage system (BESS)}

The energy balance in the BESS is as follows,
\begin{equation}
    Q^{t+\Delta t}_\text{b} = \left(\eta^+_\text{b}P^{+,t}_\text{b}-\frac{P^{-,t}_\text{b}}{\eta^-_\text{b}}\right)\Delta t+ (1-\varepsilon_\text{b}\Delta t)Q^t_\text{b}
\end{equation}
where $Q_\text{b}$ is the energy content of the BESS, $\eta^+_\text{b}$ and $\eta^-_\text{b}$ are the charge and discharge efficiencies, respectively, $P^+_\text{b}>0$ and $P^-_\text{b}>0$ are the power input and output, respectively, and $\varepsilon_\text{b}$ is the self-discharge rate. The net power to the BESS is $P_\text{b}=P_\text{b}^+-P_\text{b}^-$. The charge and discharge efficiencies and self-discharge rate of batteries are typically simplified to be independent of BESS state-of-charge and C-rates. The charge and discharge efficiencies occur both internally in the BESS and in power conversion (due to conduction and switching losses in components). The round-trip efficiency of lithium-ion BESS is typically in the range of 85\% to 95\% \cite{mongird2020energy}, meaning charge and discharge efficiencies are in the range of approximately 92\% to 97\%. The self-discharge in lithium-ion batteries are in general very low, about 3\%/per month \cite{zimmerman2004self, mongird2020energy}.

This section has given a simplified mathematical description of electrode boiler, steam accumulator, and BESS. In the following section we will use the derived equations to specify the system constraints for each component of the system. 

\section{Methodology}

\subsection{System constraints}
The system constraints are as follows. The steam demand of the plant must be met by steam produced from the electrode boiler (eb), which can be stored in the steam accumulator (sa),
\begin{equation}
    \dot{m}^t_\text{eb}+\dot{m}^t_\text{sa} = \dot{m}^t_\text{plant}.
\end{equation}
The superscript $t$ represents the time point $t$ in the analysis. As stated in section \ref{sec:system:steam}, the steam mass production rates are proportional to the electric power input,
\begin{equation}
    \dot{m}^t_\text{eb} = \frac{P^t_\text{eb}}{\Delta h_\text{tot}},
\end{equation}
where $P^t_\text{eb}$ is the electrical power inputs to the electrode boiler, $\Delta h_\text{tot}$ is the enthalpy needed to produce the steam from the inlet temperature and pressure, $T_0, p$. We will investigate the effect of preheating, \textit{i.e.} advantages of increasing $T_0$. The power to the electrode boiler is limited by its maximum power capacity,
\begin{equation}
    0 \leq P^t_\text{eb} \leq P_\text{eb}^{\max}.
\end{equation}
The mass balance in the steam accumulator is
\begin{equation}
    M^{t+\Delta t} = M^t_\text{sa} + \dot{m}^t\Delta t + \dot{m}_\text{loss},
\end{equation}
where $\dot{m}_\text{loss}$ is the the steam mass loss due to heat loss in the piping and the tank itself, as described in section \ref{sec:system:steam}. The mass storage capacity of the electrode boiler and steam accumulator, are constrained by
\begin{equation}
    0 \leq M^t_\text{sa} \leq M_\text{sa}^{\max},
\end{equation}
where $M_i^{\max}$ is total mass capacity in the various units.

The electric power balance is
\begin{equation}
    P^t_\text{grid} = P^t_\text{eb} +P^t_\text{b},
\end{equation}
where the electric power demand for the electrode boiler (eb) and BESS (b) must be met by the electric power grid. The energy balance in the BESS is
\begin{equation}
    Q^{t+\Delta t}_\text{b} = \left(\eta^+_\text{b}P^{+,t}_\text{b}-\frac{P^{-,t}_\text{b}}{\eta^-_\text{b}}\right)\Delta t+ (1-\varepsilon_\text{b}\Delta t)Q^t_\text{b},
\end{equation}
where $Q^t_\text{b}$ is the energy stored in the BESS at time point $t$, $P^{+,t}_\text{b}$ is the charging power and $P^{-,t}_\text{b}$ is the discharging power to the BESS. The charging and discharging power to the BESS is constrained by
\begin{equation}
    P_\text{b}^t = P_\text{b}^{+,t}-P_\text{b}^{-,t}, \quad P_\text{b}^{+,t} \geq 0,\quad\text{and}\quad P_\text{b}^{-,t} \geq 0
\end{equation}
The charge and discharge efficiencies are $\eta^+_\text{b}$ and $\eta^-_\text{b}$, respectively, and we have assumed them to be equal. The self-discharge rate is $\varepsilon_\text{b}$. The energy and power of the BESS is constrained by
\begin{equation}
    0.1Q_\text{b}^{\max} \leq Q^t_\text{b} \leq 0.9Q_\text{b}^{\max} \quad\text{and}\quad -\gamma Q_\text{b}^{\max} \leq P^t_\text{b} \leq \gamma Q_\text{b}^{\max},
\end{equation}
where $Q_\text{b}^{\max}$ is the energy storage capacity of the BESS and $\gamma$ is the C-rate (charge/discharge rate) of the BESS. The BESS is constrained to stay between 10 and 90\% state-of-charge, which is a typical constraint from the BESS management system to reduce BESS degradation. This effectively means that the BESS has 80\% of its stated capacity.

The participation in the FCR market requires a commitment of power capacity to up-regulation (reducing power consumption from the grid) and down-regulation (increasing power consumption from the grid). The available power capacity for FCR bids is limited by the system's current operation and maximum capacity constraints.

For down-regulation, the available capacity is limited by the remaining unused power capacity of the electrode boiler and BESS
\begin{equation}
    0 \leq P_\text{fcr}^t \leq (P_\text{eb}^{\max}-P_\text{eb}^t) + (\gamma Q_\text{b}^{\max}-P_\text{b}^t).
\end{equation}
For up-regulation, the available capacity is limited by the current total power consumption
\begin{equation}
    P_\text{fcr}^t \leq P_\text{eb}^t + P_\text{b}^t.
\end{equation}
For example, if a \SI{3}{\mega\watt} electrode boiler is currently operating at \SI{2}{\mega\watt}, it can bid at most \SI{1}{\mega\watt} for FCR, ensuring it has sufficient capacity to both increase consumption by \SI{1}{\mega\watt} (down-regulation) and decrease consumption by \SI{1}{\mega\watt} (up-regulation). The offered stand-by capacity for FCR is locked in for the duration of an hour,
\begin{equation}
    P_\text{fcr}^t = P_\text{fcr}^{t+\Delta t}\quad\forall t \quad \text{within an hour}
\end{equation}
Numerically, this condition is enforced by setting the time step to one hour ($\Delta t=3600s$). We will assume the activation of the stand-by capacity will have negligible influence on the operation of the energy system. This is because FCR services have a response time of 30 seconds and a duration of 15 minutes. This assumption simplifies the optimization model while maintaining sufficient accuracy for evaluating system economics and sizing. The minimum capacity requirement (\SI{0.1}{\mega\watt} in Norway and \SI{1}{\mega\watt} in Germany) in the real system has been disregarded in this analysis to mimic the aggregation of multiple assets, reflecting the collective behavior of such systems. In this work, we have assumed half of the FCR bids to be accepted, chosen randomly. 

The initial condition of the steam accumulator and BESS are set to 90\% of state-of-charge,
\begin{equation}
    M_\text{sa}^{t=0} = 0.9M_\text{sa}^{\max}\quad\text{and}\quad Q_\text{b}^{t=0} = 0.9Q_\text{b}^{\max}.
\end{equation}

\begin{table}[]
    \caption{Physical properties of the system}
    \centering
    \begin{tabular}{clll}
         \toprule
         Symbol & Description & Value & Unit  \\
         \midrule
         $\Delta h_\text{vap}$ & Total specific enthalpy at $T_0=283$K & 2772 & \si{\kilo\joule\per\kilo\gram}\\
         $T_0$ & Inlet temperature & 283 & \si{\kelvin}\\
         $T_a$ & Ambient temperature & 283 & \si{\kelvin}\\
         $L^+=L^-$ & Length of steam pipes & 300 & \si{\meter}\\
         $r$ & Radius of steam pipes & 10 &\si{\centi\meter}\\
         $\lambda^\text{pipe}$ & Thermal conductivity of pipe wall and insulation&0.1&\si{\watt\per\meter\per\kelvin}\\
         $\lambda^\text{tank}$ & Thermal conductivity of tank wall and insulation&0.1&\si{\watt\per\meter\per\kelvin}\\
         $\delta^\text{pipe}$ & Thickness of pipe wall and insulation&4&\si{\centi\meter}\\
         $\delta^\text{tank}$ & Thickness of tank wall and insulation&20&\si{\centi\meter}\\
         $\eta_\text{b}^+$=$\eta_\text{b}^-$& Charge and discharge efficiencies of BESS& 95 & \si{\percent}\\
         $\varepsilon_\text{b}$ &Self-discharge rate of BESS & 3 & \si{\percent\per\month}\\
         \bottomrule
    \end{tabular}
    \label{tab:physical_properties}
\end{table}

A list of physical properties of the energy network is given in table~\ref{tab:physical_properties}. For the given properties of the steam network pipes and tank, the self-discharge rate is \SI{13.3}{\percent\per\month} for a \SI{1000}{\kilo\gram} steam accumulator at maximum capacity and the charge and discharge efficiencies are \SI{90.8}{\percent} for a \SI{1}{\mega\watt} electrode boiler operating at maximum capacity. In comparison, the charge and discharge efficiencies of BESS are \SI{95}{\percent} and the self-discharge of BESS is \SI{3}{\percent\per\month} The total specific enthalpy listed in this table is with an inlet temperature of \SI{283}{\kelvin}. With an inlet temperature of \SI{366}{\kelvin} the total specific enthalpy is reduced to \SI{2425}{\kilo\joule\per\kilo\gram}.

\subsection{Energy costs}

The electricity cost is the sum of the spot price of electricity and the grid tariffs. The spot price is an hourly price per amount of energy traded in an open electricity market. The grid tariff is the sum of the volumetric and capacity components. The volumetric component is a fixed hourly price per amount of energy, while the capacity component is a function of the maximum power drawn from the grid in a given month. The total electricity cost is then
\begin{equation}
    C_\text{E} = C_\text{S} + C_\text{G}+C_0
\end{equation}
where $C_\text{S}$ and $C_\text{G}$ are the total spot price and grid tariff costs. The spot price cost is equal to
\begin{equation}
    \label{eq:spot_price_cost}
    C_\text{S} = \Delta t\sum_{t=1}^Tc_\text{s}^tP_\text{grid}^t
\end{equation}
where $\Delta t$ is the time step, $c_\text{s}^t$ is the hourly spot price at time $t$, and $T$ is the number of time step of the simulation. This is also valid when power is delivered to the grid, $P_\text{grid}^t < 0$, in which case it will act as profits for selling to the grid. We assume there to be no fees for providing energy to the grid. The grid tariff cost is the sum of the volumetric and capacity components,
\begin{equation}
    C_\text{G} = C_\text{ec} + C_\text{pc}.
\end{equation}
The volumetric component of the grid tariff cost is
\begin{equation}
    C_\text{ec} =
    \begin{cases}
    \Delta t\sum\limits_{t=1}^{T} c_\text{ec}^t P_\text{grid}^t & \text{if } P_\text{grid}^t \geq 0, \\
    0 & \text{if } P_\text{grid}^t < 0.
    \end{cases}
\end{equation}
where $c_\text{ec}^t$ is the volumetric component price. The capacity component is
\begin{equation}
    C_\text{pc} =
    \begin{cases}   
    \ c_\text{pc}\max(P_\text{grid}^t) & \text{if }\max(P_\text{grid}^t) \geq 0, \\
    0 & \text{if } \max(P_\text{grid}^t) < 0.
    \end{cases}
\end{equation}
$c_\text{pc}$ is the capacity component price and $\max(P_\text{grid}^t)$ is the maximum hourly power drawn from the grid for the period month. Typically, this maximum hourly power consumption is calculated as the mean of the three maximum power consumptions during a month. However, in this work we will simplify to consider the maximum in the whole simulation period of one year. 

When purchasing electricity from the grid, one pays the spot price, the grid tariffs, and other taxes and fees. However, when selling electricity back to the grid, the profit is determined solely by the spot price at that time. This means that the difference in the spot price must account for the grid tariffs and charge-discharge efficiencies in order for price arbitrage to be profitable. 

Spot prices in the day-ahead market are established one day in advance, typically varying on an hourly basis throughout the delivery day, and are set by the dynamics of the open electricity market. The grid tariffs are determined by the grid operator. In this work we will consider historical data from the period of 1st of January 2024 to 1st of January 2025.

\begin{table}
    \centering
    \caption{Summary of electricity costs and profits.}
    \label{tab:costs}
    \begin{tabular}{cll}
        \toprule
        Symbol & Description & Unit\\
        \midrule
        $c_\text{s}^t$ & Hourly electricity spot price & \si{EUR\per\kilo\Wh}\\
        $c_\text{ec}^t$ & Hourly electricity grid tariff volumetric component &\si{EUR\per\kilo\Wh}\\
        $c_\text{pc}$ & Electricity grid tariff capacity component & \si{EUR\per\kilo\watt\per\month}\\
        $p_\text{fcr}^t$ & Capacity price for FCR participation & \si{EUR\per\kilo\watt}\\
        \bottomrule
    \end{tabular}
\end{table}

The initial energy cost $C_0$ is taken into account, which is the cost of charging the steam accumulator and BESS up to their initial conditions,
\begin{equation}
    C_0 = \overline{(c_\text{s}+c_\text{ce})}\left[Q_\text{b}^{t=0}+M_\text{sa}^{t=0}\Delta h_\text{tot}\right]
\end{equation}
where superscript $t=0$ indicates the value at time $t=0$, meaning the initial condition. The cost associated to the initial conditions is taken from the sum of the mean spot price and the mean volumetric component of the grid tariff for the simulation period, calculated as 
\begin{equation}
    \overline{(c_\text{s}+c_\text{ce})} = \frac{1}{T}\sum_{t=1}^T(c_\text{s}^t+c_\text{ce}^t).
\end{equation}
The capacity component of the grid tariff is excluded from this initial cost calculation, based on the assumption that the power required to charge the energy storages to their initial state is less than the maximum power drawn during the rest of the time period. However, for very large energy storages, this assumption may not hold and will be verified prior to analysis.

\subsection{Profits from ancillary services and price arbitrage}

Profits can be obtained from price arbitrage and from providing ancillary services. This is possible in this system because there is energy storage capacity in the lithium-ion batteries. Profits from price arbitrage are contained in the total spot price cost, $C_\text{S}$. Ancillary services, in this case FCR services, are provided by asset owners to the grid operators in order to stabilize the electricity grid. The profits from providing FCR services is
\begin{equation}
    \Pi_\text{fcr} = \frac{\Delta t}{\SI{1}{\hour}}\sum_{t=1}^Tp^t_\text{fcr}P^t_\text{fcr}
\end{equation}
where $p_\text{fcr}^t$ is the hourly FCR price for stand-by and $P_\text{fcr}^t$ is the hourly capacity committed to the FCR market in that hour. The time step $\Delta t$ is normalized by one hour, as the stand-by capacity is locked in for that hour, while the time step might differ from one hour. There are no profits from activation in the FCR market in Germany or Norway, only for having capacity on stand-by. In Germany, in the 50Hertz transmission region, the bid period is 15 minutes, while in Norway it is 1 hour. We will in this analysis simplify the bid period to be 1 hour for both regions.

\subsection{Formal definition of optimization problem}
\label{sec:formal}

\begin{table}[]
    \renewcommand{\arraystretch}{1.5}
    \centering
    \caption{List of variables in the optimization problem.}
    \begin{tabular}{lll}
        \toprule
        Symbol & Description & Unit \\
        \midrule
        $t$ & Point in time & \si{\second}\\
        $\Delta t$ & Time step & \si{\second}\\
        $P^t_\text{grid}$ & Power drawn from electricity grid & \si{\watt}\\
        $P^t_\text{eb}$ & Power supplied to electrode boiler & \si{\watt}\\
        $P^t_\text{b}$ & Net power supplied to BESS & \si{\watt}\\
        $P^{+,t}_\text{b}$ & Power supplied to BESS during charging & \si{\watt}\\
        $P^{-,t}_\text{b}$ & Power supplied by BESS during discharging & \si{\watt}\\
        $P^t_\text{fcr}$ & Power in stand-by to participate in FCR market & \si{\watt}\\
        $P_\text{eb}^{\max}$ & Max power capacity of electrode boiler &\si{\watt}\\
        $P_\text{b}^{\max}$ & Max power capacity of BESS &\si{\watt}\\
        $\dot{m}^t_\text{eb}$ & Steam mass flow rate from electrode boiler &\si{\kilo\gram\per\second}\\
        $\dot{m}^t_\text{sa}$ &  Steam mass flow rate to steam accumulator &\si{\kilo\gram\per\second}\\
        $\dot{m}^t_\text{plant}$ & Steam mass flow rate to plant &\si{\kilo\gram\per\second}\\
        $Q^t_\text{b}$ & State-of-charge of BESS &\si{\joule}\\
        $Q_\text{b}^{\max}$ & Maximum state-of-charge of BESS &\si{\joule}\\
        $M^t_\text{sa}$ & Steam mass content in steam accumulator &\si{\kilo\gram}\\
        $M_\text{sa}^{\max}$ & Maximum steam mass content in steam accumulator &\si{\kilo\gram}\\
        \bottomrule
    \end{tabular}
    \label{tab:variables}
\end{table}

The objective of the optimization problem is to minimize the net energy costs. A list of symbols in this optimization problem with description is given in table \ref{tab:variables}.

\textbf{Objective Function:}
\begin{equation}
    \min \quad C_\text{E}-\Pi_\text{fcr}
\end{equation}
where $\Pi_\text{fcr}$ is profits from participation in the FCR market and $C_\text{E}$ is the total electricity cost, which includes the spot price and grid tariff costs.

\textbf{Subject to the Following Constraints:}
\begin{equation}
    \begin{aligned}
    1. \quad & \dot{m}^t_\text{eb}+\dot{m}^t_\text{sa} = \dot{m}^t_\text{plant},\quad \forall t,\\
    2. \quad & \dot{m}^t_\text{eb} = \frac{P^t_\text{eb}}{\Delta h_\text{tot}},\quad \forall t,\\
    3. \quad & P_\text{grid}^t = P_\text{eb}^t + P_\text{b}^t, \quad \forall t, \\
    4. \quad & P_\text{fcr}^t \leq (P_\text{eb}^{\max} - P_\text{eb}^t) + (\gamma_\text{b} Q_\text{b}^{\max} - P_\text{b}^t), \quad \forall t,\\
    5. \quad & P_\text{fcr}^t \leq P_\text{eb}^t + P_\text{b}^t, \quad \forall t,\\
    6. \quad & P^t_\text{fcr} = P_\text{fcr}^{t+1h},\quad \forall t \text{ within an hour}\\
    7. \quad & 0 \leq M_\text{sa}^t \leq M_\text{sa}^{\max}, \quad \forall t, \\
    8. \quad & 0 \leq P_\text{eb}^t \leq P_\text{eb}^{\max}, \quad \forall t, \\
    9. \quad & 0.1Q_\text{b}^{\max} \leq Q_\text{b}^t \leq 0.9Q_\text{b}^{\max}, \quad \forall t, \\
    10. \quad & Q^{t+\Delta t}_\text{b} = \left(\eta^+_\text{b}P^{+,t}_\text{b}-\frac{P^{-,t}_\text{b}}{\eta^-_\text{b}}\right)\Delta t+ (1-\varepsilon_\text{b}\Delta t)Q^t_\text{b}, \quad \forall t,\\
    11. \quad & P_\text{b}^t = P_\text{b}^{+,t}-P_\text{b}^{-,t}, \quad \forall t,\\
    12. \quad & P_\text{b}^{+,t} \geq 0,\quad \forall t,\\
    13. \quad & P_\text{b}^{-,t} \geq 0,\quad \forall t,\\
    11. \quad &-\gamma Q_\text{b}^{\max} < P_\text{b}^t\leq\gamma Q_\text{b}^{\max},\quad \forall t\\
    12. \quad & M^{t+\Delta t} = M^t_\text{sa} + \dot{m}^t\Delta t + \dot{m}_\text{loss}, \quad \forall t,\\
    13. \quad & Q_\text{b}^{t=0} = 0.9Q_\text{b}^{\max}, \text{ and}\\
    14. \quad & M_\text{sa}^{t=0} = 0.9M_\text{sa}^{\max}.
    \end{aligned}
\end{equation}

\textbf{Decision Variables:}
\begin{equation}
    P_\text{eb}^t, P_\text{b}^{+,t}, P_\text{b}^{-,t}, P_\text{fcr}^t  \quad \forall t
\end{equation}

\subsection{Investment and maintenance cost}

\begin{table}[]
    \centering
    \caption{Power law functions are used to describe the size dependent investment cost for the electrode boiler, steam accumulator and the BESS. This table lists the coefficients used for these power laws.}
    \label{tab:investment_parameters}
    \begin{tabular}{lcccc}
    \toprule
    Description & Symbol & Electrode boiler & BESS &Steam accumulator\\
    \midrule
    Base cost coefficient &$c_i$ &\SI{152}{EUR\per\kilo\watt}& \SI{433}{EUR\per\kilo\Wh} & \SI{191}{EUR\per\kilo\gram}\\
    Capacity exponent &$\alpha_i$ &---&-0.164& -0.05\\
    Power exponent & $\beta_i$ &-0.296&0.005 & ---\\
    \bottomrule
    \end{tabular}
\end{table}

\begin{figure}
    \centering
    \begin{subfigure}[b]{0.49\textwidth}
        \centering
        \includegraphics[width=\textwidth]{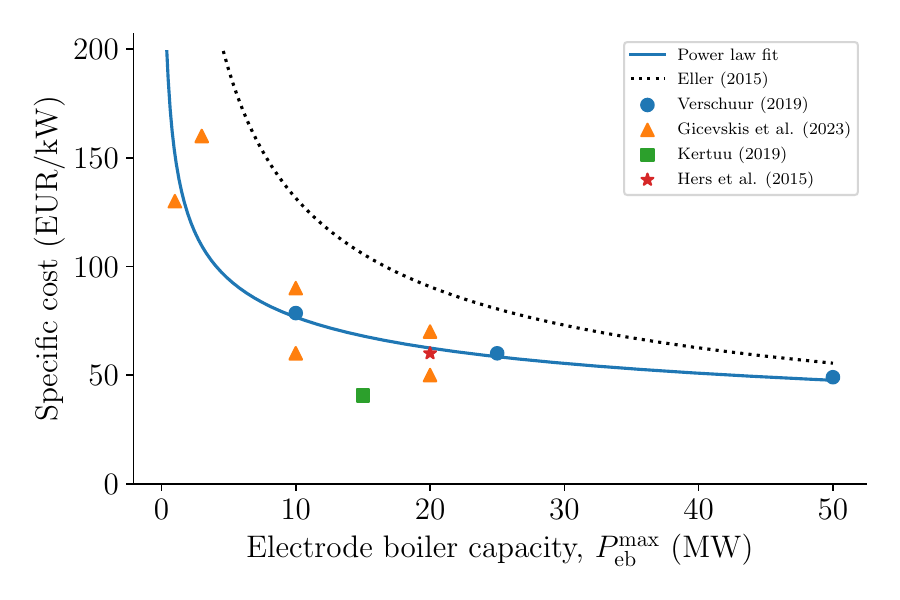}
        \caption{}
        \label{fig:investment_eb_cost}
    \end{subfigure}
    \hfill
    \begin{subfigure}[b]{0.49\textwidth}
        \centering
        \includegraphics[width=\textwidth]{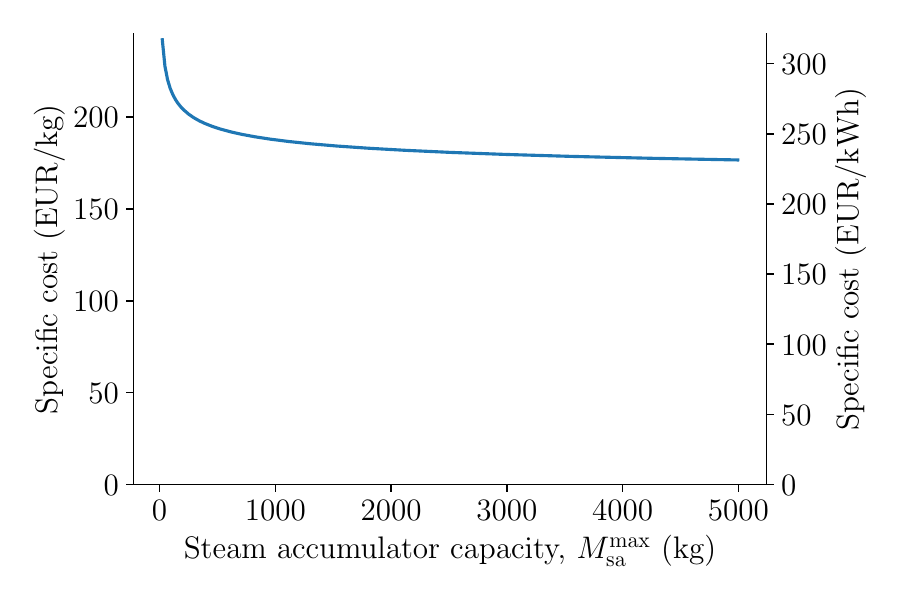}
        \caption{}
        \label{fig:investment_sa_cost}
    \end{subfigure}
    \caption{(a) Estimated specific investment costs for electrode boiler as a function of electrode boiler power capacity \cite{eller2015integration, verschuur2019exploratory, gicevskis2023role, kerttu2019evaluation, hers2015potential}. The various sources are shown as markers and the fitted power law is shown as a solid blue line. The power law of Eller \cite{eller2015integration} in shown in dashed black line. (b) Estimated specific investment costs of steam stored for steam accumulator as a function of the steam accumulator steam mass capacity \cite{beck2019kostenoptimierte}.}
    \label{fig:investment_costs_eb_sa}
\end{figure}

The investment costs for the electrode boiler and BESS are assumed to follow power law functions of the unit sizes. The individual investment costs are
\begin{equation}
    \label{eq:power_law}
    C_\text{eb} = c_\text{eb}P_\text{eb}^{\max}\left(\frac{P_\text{eb}^{\max}}{P_0}\right)^{\beta_\text{eb}},\quad C_\text{sa} = c_\text{sa}M_\text{sa}^{\max},\quad
    C_\text{b} = c_\text{b}Q_\text{b}^{\max}\left(\frac{Q_\text{b}^{\max}}{Q_0}\right)^{\alpha_\text{b}}\gamma^{\beta_\text{b}},
\end{equation}
where $P_\text{eb}^{\max}$ is the power capacity of the electrode boiler, $M_\text{sa}^{\max}$ is the mass capacity of the steam accumulator, $Q_\text{b}^{\max}$ is the energy storage capacity of the BESS and $\gamma$ is the charge and discharge rate of the BESS (C-rate). The symbols $P_0$, $M_0$, and $Q_0$ are constant reference values (i.e., \SI{1}{\mega\watt}, \SI{1e3}{\kilo\gram}, \SI{1}{\mega\Wh}) used to normalize the respective quantities in parentheses, ensuring dimensionless expressions. These equations reflect the principle of economy of scale, where it is more cost-efficient to buy larger units. A similar power law is given by Eller for the cost of electrode boilers \cite{eller2015integration}. This is often because of an initial installation cost that is independent of unit size, or that there are components of the unit that does not depend on the unit size. For example, larger BESS may require only incremental increase in power electronics, cooling systems, and electronics for battery management systems. 

\begin{figure}
    \centering
    \includegraphics[width=0.75\textwidth]{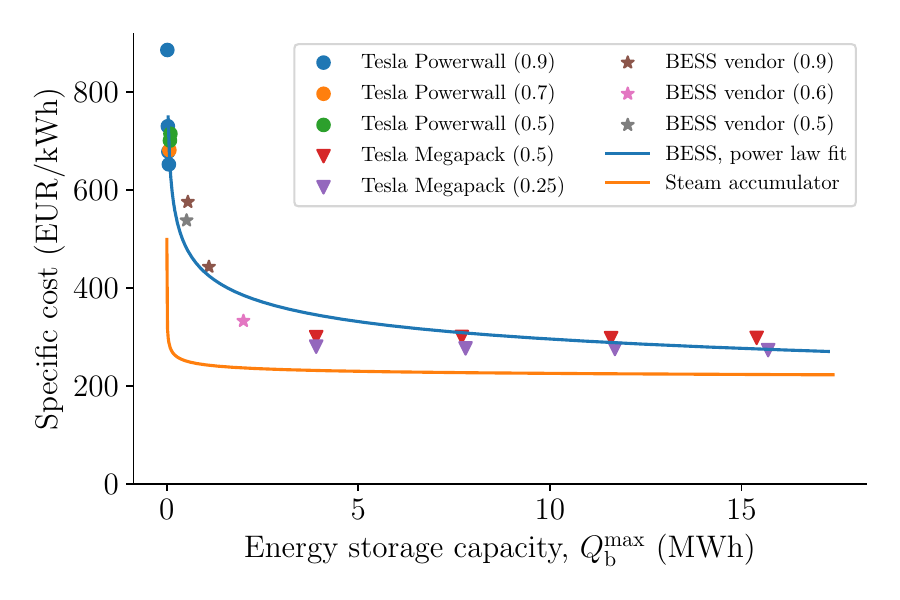}
    \caption{Estimated specific investment costs for a BESS, based on Tesla Powerwall, Megapack, and an unnamed vendor \cite{TeslaMegapackDesign, TeslaPowerwallDesign}. The C-rates of the various BESS' are given parentheses in the legend. The specific investment cost for a steam accumulator is shown in orange for reference.}
    \label{fig:investment_b_cost}
\end{figure}

Various sources have been used to determine the coefficients of the power law functions. This is presented in figure~\ref{fig:investment_eb_cost} for the electrode boiler, figure~\ref{fig:investment_sa_cost} for the steam accumulator, and figure~\ref{fig:investment_b_cost} for the BESS. The scaling coefficients $c_i$, capacity exponents $\alpha_i$, and power exponents $\beta_i$, are given in table~\ref{tab:investment_parameters}. The power law used in this analysis for the electrode boiler has been fitted to data from sources in academic literature \cite{verschuur2019exploratory, gicevskis2023role, kerttu2019evaluation, hers2015potential}. Although these sources span from 2015 to 2023 and may be somewhat outdated, the cost of electrode boilers is largely driven by fluctuations in raw material prices, given that the underlying technology is mature and well-established. In contrast, the cost of battery cells has dramatically decreased in recent years, and is expected to continue decreasing. As a result, the cost of utility scale BESS has also decreased significantly. The BESS cost data used in this analysis is based on information from unnamed utility-scale BESS vendor in Norway ("Vendor A") and the public available pricing for the Tesla Powerwall and Megapack \cite{TeslaPowerwallDesign, TeslaMegapackDesign}, all with delivery dates in 2025. An installation cost of 18\% has been assumed for Tesla Megapack, and 25\% for Tesla Powerwall. This is based on an installation cost varying between 18\% and 23\% from the unnamed vendor. For steam accumulators, the cost function by Beck and Drexler-Schmid is followed \cite{beck2019kostenoptimierte}, with an added 20\% installation and auxiliary utilities (sensors, valves, etc) cost. This function assumes the steel of pressure vessel of the steam accumulator to be main contributor to its cost, with an assumed steel cost of \SI{4}{EUR\per\kilo\gram} and maximum pressure capacity of \SI{15}{\bar}. This cost function has been fit to the power law described in equation \ref{eq:power_law}.

The total investment cost is
\begin{equation}
    C_\text{I} = C_\text{sb}+C_\text{sa}+C_\text{b}.
\end{equation}
and the annual maintenance costs are assumed to be 2\% of the investment costs, $C_\text{M}=0.02C_\text{I}$. The net present value of the system is
\begin{equation}
    \text{NPV} = \sum_{t=0}^{L} \frac{\Pi_\text{fcr} - C_\text{E} - C_\text{M}}{(1 + r)^t} - C_\text{I}
\end{equation}
where $\Pi_\text{fcr}$ is the profits from ancillary market participation, $C_\text{E}$ is the total electricity costs, $C_\text{M}$ is the maintenance costs, $C_I$ is the investment cost, $r$ is the depreciation rate, set to 5\%, and $L$ is the lifetime of the equipment, assumed to be 15 years. We will use the relative net present value, $\Delta \text{NPV} = \text{NPV} - \text{NPV}_0$, where the reference $\text{NPV}_0$ is the NPV without storage installed. In this work, we estimate the net energy costs for the year of 2024, and we extrapolate this through the whole 15-year lifetime of the equipment. This is a simplification, and it is highly likely that the conditions, such as energy prices and FCR market profits, will change during the 15-year lifetime of the equipment.

A grid search is performed over a range of BESS capacities and C-rates, steam accumulator capacities, and electrode boiler capacities. For each combination, a single-period optimization is performed, minimizing the net energy cost ($C_\text{E}-\Pi_\text{fcr}$) as defined in section \ref{sec:formal}, for the period from 1st of January 2024 to 1st of January 2025. This single-period optimization is implemented using CVXPY (version 1.5.2) \cite{diamond2016cvxpy, agrawal2018rewriting}. The results from this single-period optimization, along with power-law cost functions for the equipment, are used to calculate the NPV over the assumed 15-year lifetime of the system. The grid search results identify promising regions of the parameter space. To refine the solution and find a more precise optimum, a second-stage optimization is performed using the differential evolution algorithm \cite{Storn1997}, as implemented in the SciPy library (version 1.10.1) \cite{2020SciPy-NMeth, scipy_differential_evolution}. The decision variables for this stage are the BESS capacity and C-rate, steam accumulator capacity, and electrode boiler capacity. This two-stage optimization was selected to efficiently explore the broad parameter space (grid search), followed by precise local optimization (differential evolution), ensuring robustness while managing computational complexity. 

\subsection{Model limitations and assumptions}
Several simplifying assumptions were made in this study to maintain computational feasibility and clarity of the results. First, perfect foresight was assumed for steam demand, electricity spot prices, and FCR market prices, which is an idealization that can lead to overestimating potential profits compared to real operational conditions with forecast errors. Second, the dynamic effects of FCR activation on system operation were neglected, assuming that the relatively short activation duration (15 minutes) and limited response frequency have negligible impact on overall energy balances and constraints. Third, battery cell degradation, including loss of capacity and efficiency over time, was not accounted for, potentially leading to optimistic long-term economic evaluations of BESS. We are however operating the BESS with state-of-charges between 10\% and 90\% and at relatively low C-rates (below 0.9), which should reduce battery cell degradation. Additionally, thermodynamic and fluid dynamic processes were simplified, assuming constant pressures, temperatures, and perfect phase transitions. Lastly, minimum bid requirements for FCR market participation were disregarded, implicitly assuming aggregated bidding from multiple smaller assets.

\section{Input data}

Historical data has been sourced for the steam demand, electricity spot prices, grid tariffs, and FCR market prices. The time period of 1st of January 2024 to 1st of January 2025 has been chosen for this analysis, and two regions have been investigated, 50Hertz transmission region in Germany and electricity price area NO3 in Norway. The 50Hertz transmission region in Germany primarily covers the eastern part of the country, including Berlin and Brandenburg, while the electicity price area NO3 covers central Norway, including Trondheim and Molde. For now on we will refer to these regions as just Norway (NO) and Germany (DE). A conversion rate of 0.086 was used to convert Norwegian kroner (NOK) to euros (EUR).

\begin{figure}
    \includegraphics[width=\linewidth]{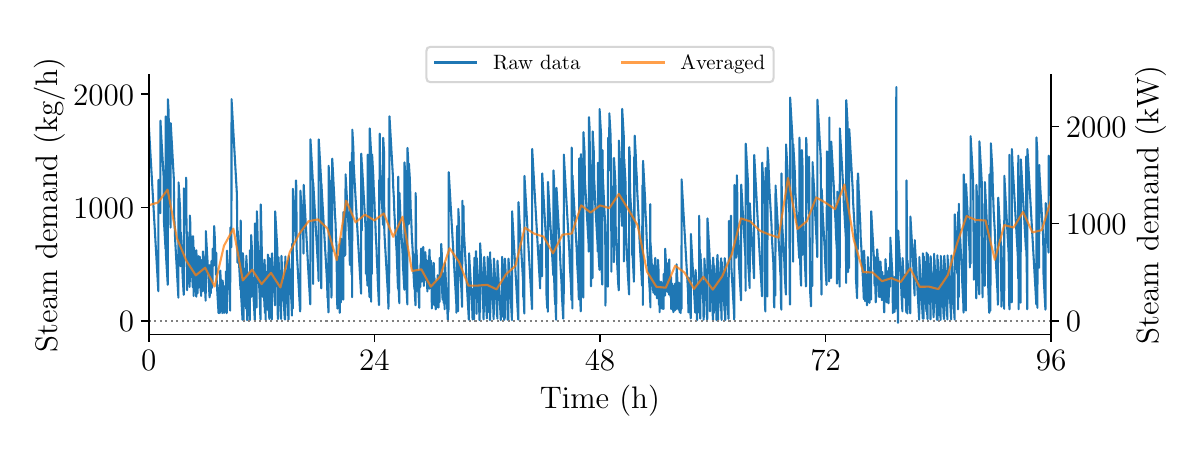}
    \caption{Raw steam demand data from a steam consuming factory. In blue the raw data with a time resolution of 1 minute, and averaged data in orange with a time resolution of 5 minutes.}
    \label{fig:raw_steam}
\end{figure}

\begin{figure}
    \includegraphics[width=\linewidth]{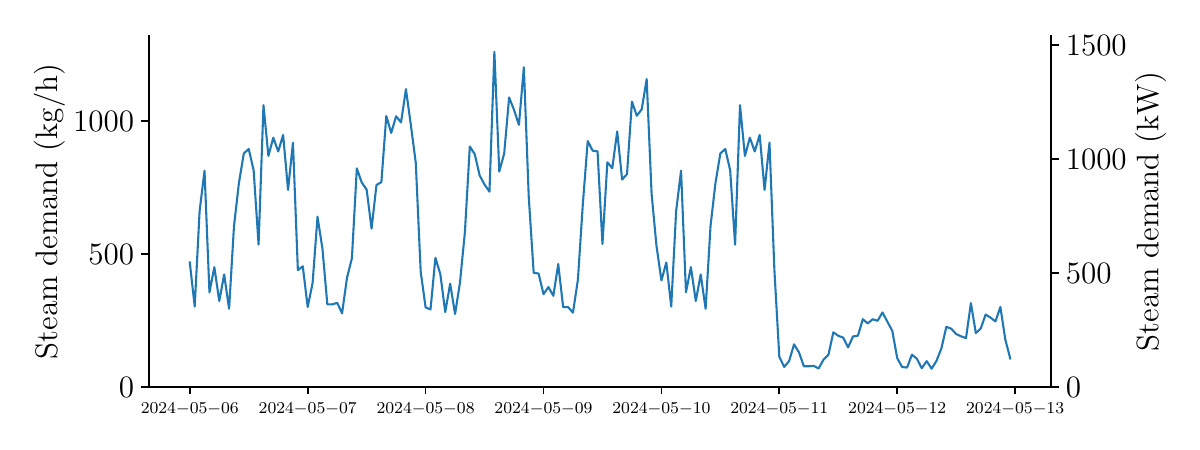}
    \caption{Steam demand data from a steam consuming factory, here shown for a week from Monday through Sunday. The raw data has been repeated periodically to extend the dataset to 8784 hours (366 days), and averaged to give a time resolution of 1 hour.}
    \label{fig:steam_demand}
\end{figure}

The original steam demand data was for a period of 96 hours (4 days) and a time resolution of 1 minute, presented in figure~\ref{fig:raw_steam}. The raw steam data presented here was provided by a steam consuming industry site, and starts at 11.00 on a Monday (hour 0), and ends at 11.00 on a Friday (hour 96). This original data has been repeated periodically to extend the dataset to 8784 hours (366 days, 2024 was a leap year), and averaged to give a time resolution of 1 hour. This data is presented for a week in May (6th to 13th of May 2024) in figure~\ref{fig:steam_demand}. Averaging was applied to improve the computational efficiency of the optimization procedure, and the resulting data was repeated periodically to introduce greater variability in the spot price and FCR capacity price. As a side effect of the averaging procedure, the peaks and valleys in the steam demand profile are smoothed out. For simplicity, we assume that such fluctuations are absorbed by the steam already present in the steam network. In addition, the steam demand has been scaled down to 25\% during Saturdays and Sundays.This gives additional opportunities for energy storage to charge up during the weekend.

\begin{figure}
    \centering
    \includegraphics[width=\linewidth]{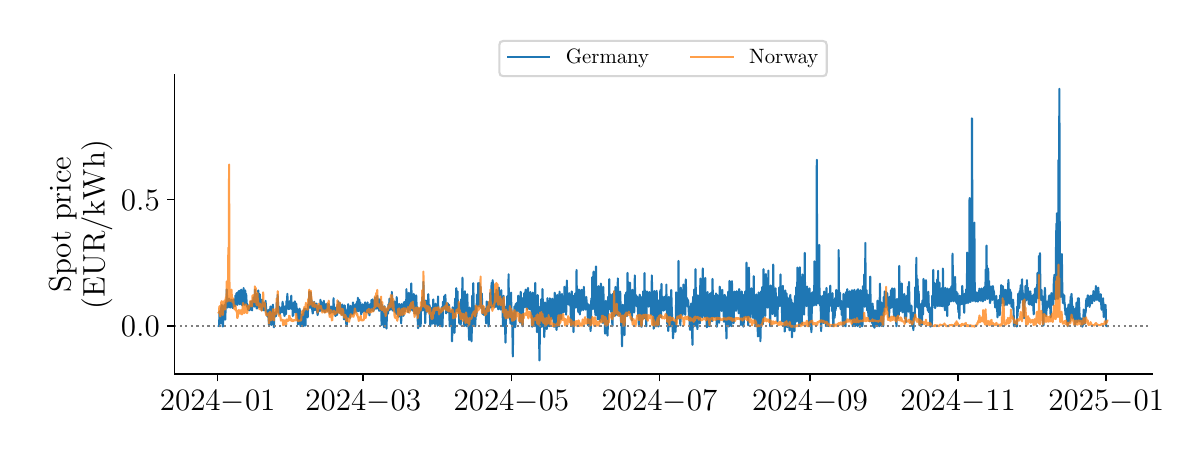}
    \caption{Spot price in the electricity price area of Germany and Norway for the period from 1st of January 2024 to 1st of January 2025.}
    \label{fig:price}
\end{figure}

The spot price data for Norway is sourced from Nord Pool \cite{nordpool}. The spot price data for Germany has been sourced from the SMARD data portal by the Federal Network Agency in Germany (Bundesnetzagentur) \cite{smard_download_center}. The spot prices are presented in figure~\ref{fig:price}. The average spot price during this period was \SI{0.0350\pm0.0003}{EUR\per\kilo\Wh} in Norway and \SI{0.0785\pm0.0006}{EUR\per\kilo\Wh} in Germany. On average, the spot price in Germany was approximately 2.25 times higher than in Norway.

The grid tariffs in Norway are sourced from Tensio, which is the distribution system operator in the NO3 region \cite{tensio}. The grid tariffs in Germany is sourced from 50Hertz, who is the transmission system operator in north eastern Germany \cite{netztransparenz_grid_fees}. The grid tariffs are sourced for high-voltage and high energy consumption businesses. The volumetric and capacity components of  the grid tariff, are presented in table~\ref{tab:grid_tariff}. The capacity component is more than 7 times higher in Germany than in Norway; however, the volumetric component  is almost 5 times higher in Norway than in Germany. It is expected that this will dramatically affect the optimal size and operation of the electrode boiler, BESS, and steam accumulator in the two countries.

\begin{table}[]
    \centering
    \caption{Components of the grid tariffs in Norway \cite{tensio} and Germany \cite{netztransparenz_grid_fees} for high-voltage grid connection.}
    \label{tab:grid_tariff}
    \begin{tabular}{lccl}
        \toprule
        Grid tariff component & Norway & Germany & Unit\\
        \midrule
        Capacity component & 4.386 & 32.11 & \si{EUR\per\kilo\watt\per\month}\\
        Volumetric component & 0.03612 & 0.0074 & \si{EUR\per\kilo\Wh}\\
        \bottomrule
    \end{tabular}
\end{table}

\begin{figure}
    \centering
    \includegraphics[width=\linewidth]{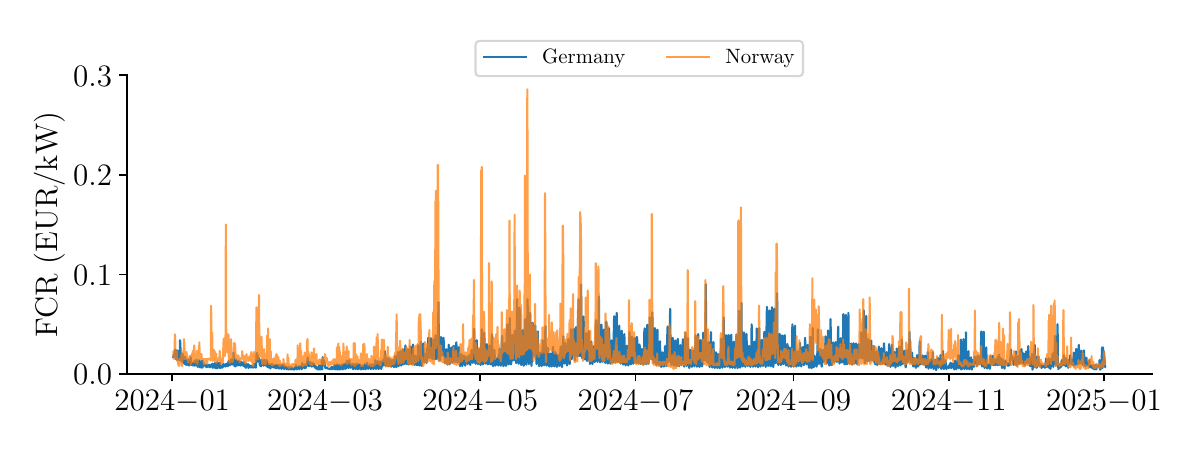}
    \caption{Capacity price for FCR from 1st of January 2024 to 1st of January 2025 for Germany (50Hertz) \cite{entso-e} and Norway (NO3) \cite{statnett_reservemarkeder}.}
    \label{fig:fcr_price}
\end{figure}

The capacity price for FCR market participation has been sourced from Statnett \cite{statnett_reservemarkeder} for Norway and from ENTSO-E Transparency Platform \cite{entso-e} for Germany, and is presented in figure~\ref{fig:fcr_price}. The average FCR capacity price in Norway for the period was \SI{0.0201\pm0.0002}{EUR\per\kilo\watt}, while in Germany it was \SI{0.01626\pm0.00012}{EUR\per\kilo\watt}. In Norway the price was approximately 1.24 higher than in Germany. The bid acceptance has been assumed to be 50\%, and the hours with accepted bids has been chosen randomly. This has been done numerically by setting the capacity price for FCR market participation to zero for 50\% of the hours in the year.

\section{Results and discussion}

\subsection{Optimal system configuration}
Table~\ref{tab:optimal_configuration} presents the most cost-efficient system configuration. The size of BESS, steam accumulator, and electrode boiler for the two electricity regions, electricity area NO3 in Norway and 50Hertz transmission region in Germany, for the period from 1st of January 2024 to 1st of January 2025 is presented. The most cost-efficient configurations without storage are also presented for reference.

\begin{table}[]
    \centering
    \caption{Optimal configuration of steam accumulator, electrode boiler, and BESS in Norway (NO) and Germany (DE). The case of no storage is also shown as references for each region.}
    \begin{tabular}{llccccl}
        \toprule
        Description & Symbol & Ref. NO & NO & Ref. DE & DE & Unit \\
        \midrule
         Steam accumulator mass capacity &  $M_\text{sa}^{\max}$ & 0 & 438 & 0& 2125 & \si{\kilo\gram}\\
         Electrode boiler capacity & $P_\text{eb}^{\max}$ & 1644 & 1702 & 1608 & 1413 &\si{\kilo\watt}\\
         BESS capacity & $Q_\text{b}^{\max}$ & 0 & 0 & 0& 0 &\si{\kilo\Wh}\\
         BESS C-rate & $\gamma$ &---& --- & --- & --- & ---\\
        \midrule
         Steam accumulator cost &  $C_\text{sa}$ & 0 & 87 & 0 & 390 & kEUR\\
         Electrode boiler cost & $C_\text{eb}$ &215 & 220 & 212 & 193 &kEUR\\
         BESS cost & $C_\text{b}$ &0 & 0 & 0 & 0 & kEUR\\
         \midrule
         \textbf{Total investment cost} & $\mathbf{C_\text{I}}$ & \textbf{215} & \textbf{308} & \textbf{212} & \textbf{583} & \textbf{kEUR}\\
         \bottomrule
    \end{tabular}
    \label{tab:optimal_configuration}
\end{table}

\begin{figure}
    \centering
    \includegraphics[width=0.9\linewidth]{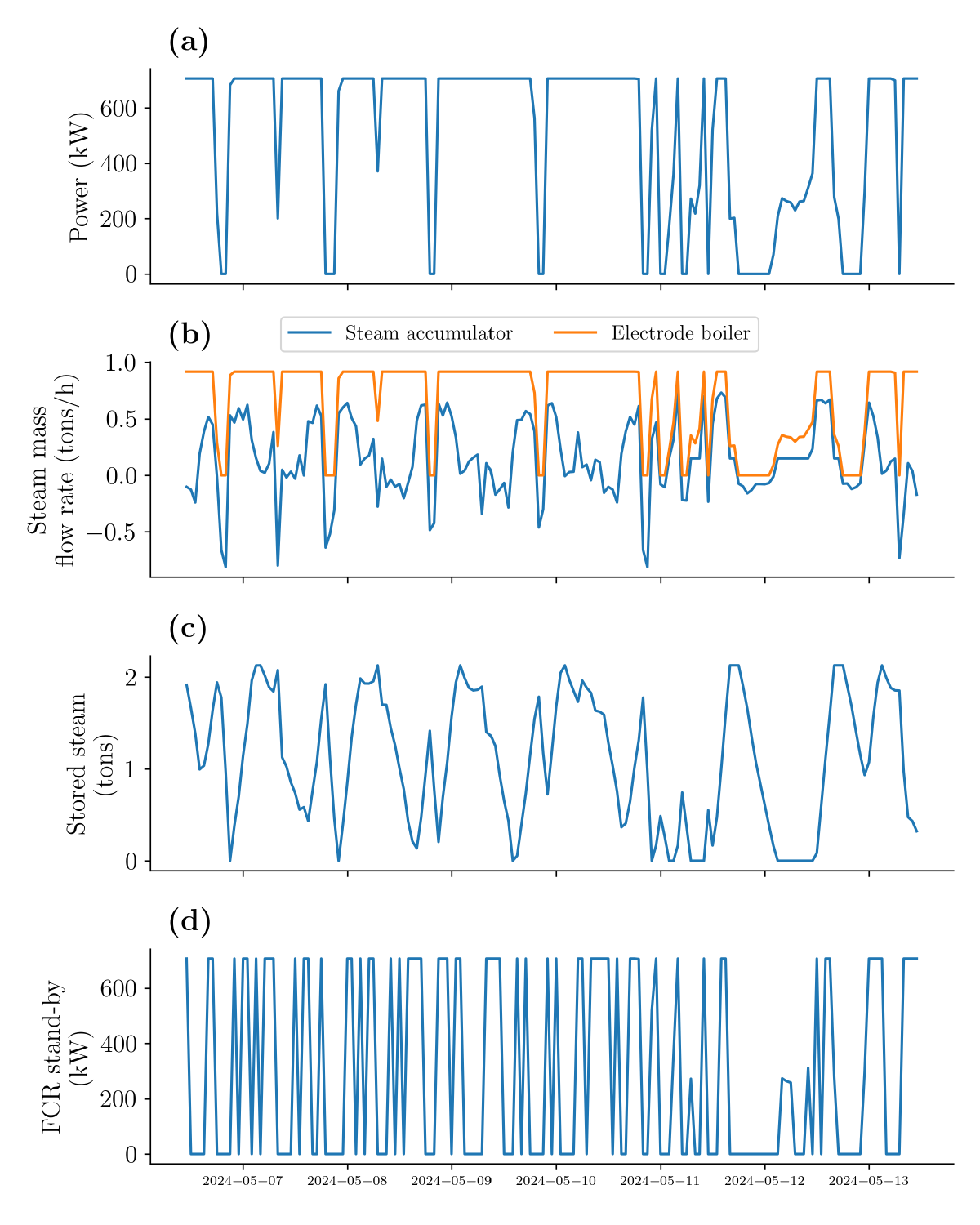}
    \caption{Time series data for a factory in Germany showing: (a) electric power to electrode boiler, (b) steam mass flow rate from electrode boiler and steam accumulator, (c) stored steam in the steam accumulator, and (d) power stand-by for FCR. The time period shown is from 11:00 on May 5th to 10:00 on May 13th 2024, for a BESS capacity of \SI{0}{\kilo\Wh}, steam accumulator mass capacity of \SI{2090}{\kilo\gram}, and electrode boiler capacity of \SI{1465}{\kilo\watt}. This is the configuration giving the optimal NPV.}
    \label{fig:DE_a_week}
\end{figure}

To illustrate the results from the optimization, a time series of decision variables that minimize the net energy costs ($C_\text{E}-\Pi_\text{fcr}$) is shown in figure~\ref{fig:DE_a_week}. The results are shown for the optimal capacities of electrode boiler, steam accumulator, and BESS in Germany, which is presented in table~\ref{tab:optimal_configuration}. The time period shown is from 11:00 on May 5th to 10:00 on May 13th, 2024. For the rest of this work, the period from 1st of January 2024 to 1st of January 2025 is considered. This figure illustrates the dynamic use of the electrode boiler and steam accumulator to meet the steam demand, while minimizing the net energy cost. 

The differences in the optimal sizing of different components between Norway and Germany emphasize how the electricity market and the tariff design fundamentally shape the economic feasibility of energy flexibility technologies. In Germany, the capacity component of the grid tariff is large compared to the volumetric component, while in Norway the opposite is true. In addition, the average spot price in Germany is approximately 3 times higher than in Norway. Both of these effects give a larger incentive to invest in energy storage in Germany than in Norway. In these simulations we find that in Germany the optimal storage capacity of the steam accumulator is \SI{2090}{\kilogram}, approximately 4.8 times more than the optimal capacity in Norway. We find that for both Norway and Germany, the investment cost of BESS is too high to prioritize this over steam accumulators. The optimization results suggest that under current cost structures, steam accumulators are consistently favored. However, as discussed in the introduction, the cost of BESS is expected to continue to reduce, while the same is not expected for steam accumulators nor electrode boilers. This in mind, we will investigate the sensitivity of the optimal configuration to the cost of electrode boilers and BESS in section \ref{sec:sensitivity}. 

\subsection{Annual costs and profits}

The total annual cost for the most cost-efficient system configurations for Norway and Germany is presented in table~\ref{tab:table_optimal}. These results show that the net energy cost ($C_\text{E}-\Pi_\text{fcr}$) is approximately two times higher in Germany than in Norway, and the largest contribution to this difference is the capacity component of the grid tariff. Our findings indicate that participating in the FCR market can be financially beneficial, yielding approximately 17\% and 7\% reduction of the net energy cost for Norway and Germany, respectively. These profits stem solely from the participation with the electrode boiler, as no BESS capacity is installed in these cases. However, the declining cost of BESS and the growing share of intermittent renewable energy are expected to impact future profitability in ancillary service markets. The net effect on future profitability remains uncertain at this stage. 

\begin{table}[]
    \centering
    \caption{Annual cost and profits associated to the spot market cost $C_\text{s}$, volumetric component of the grid tariff cost $C_\text{ec}$, capacity component of the grid tariff cost$C_\text{pc}$, and the profits from participation in FCR market $\Pi_\text{fcr}$ for the most cost-efficient configurations of steam accumulator, electrode boiler, and BESS in Norway (NO) and Germany (DE). The case of no storage is also shown as references for each region.}
    \begin{tabular}{llccccl}
        \toprule
        Description & Symbol & Ref. NO & NO & Ref. DE & DE & Unit \\
        \midrule
         Spot-market cost &  $C_\text{s}$ &158& 158 & 362 &348 & kEUR\\
         Grid tariff cost (volumetric) & $C_\text{ec}$ &158& 160 & 32 & 34 &kEUR\\
         Grid tariff cost (capacity) & $C_\text{pc}$ &56& 45 & 410 & 273 & kEUR \\
         Profits from FCR market & $\Pi_\text{fcr}$ &40& 54 & 33 & 43 & kEUR \\
        \midrule
         \textbf{Net energy cost} & $\mathbf{C_\text{E}-\Pi_\text{fcr}}$ &\textbf{331}& \textbf{310} & \textbf{772} &\textbf{612} & \textbf{kEUR}\\
        \midrule
         Annual savings potential & &---& 21 & --- & 160 & kEUR\\
         \bottomrule
    \end{tabular}
    \label{tab:table_optimal}
\end{table}

In table~\ref{tab:table_optimal_energy_power}, the power and energy consumption is shown. The analysis demonstrates that high capacity component price in Germany create a strong incentive to reduce peak loads, showing that market structure and grid pricing greatly influence the system design. Installation of a steam accumulator reduces the peak power demand by 20\% and 34\% in Norway and Germany, respectively, with the steam accumulator capacities given in table~\ref{tab:optimal_configuration}. This, however, comes at a cost of increased energy consumption of 1.4\% and 4.6\% in Norway and Germany, respectively, due to losses associated with charge, discharge and self-discharge of the steam accumulator. 

In Norway, the average number of daily cycles is 6.5, corresponding to a cycle period of approximately 3.7 hours. The steam accumulator is thus applied to buffer short-term mismatches between boiler output and demand, rather than to shift loads over longer periods. The average number of daily full steam accumulator cycles in Germany is 1.7, indicating use with a typical period of approximately 14 hours. This reflects a strategy focused on daily load shifting resulting from the higher capacity component grid tariff in Germany, which encourage for a more conservative operation with fewer, longer cycles aligned with major demand peaks.

\begin{table}[]
    \centering
    \caption{Annual energy and power consumption in Norway and Germany for the most cost-efficient configurations of a steam accumulator (SA), electrode boiler, and BESS. The case of no storage is also shown as references for each region.}
    \begin{tabular}{llccccl}
        \toprule
        Description & Symbol & Ref. NO & NO & Ref. DE & DE & Unit \\
        \midrule
         Max grid power & $\max({P_\text{grid}})$& 1063 & 851 & 1063 & 706 & \si{\kilo\watt}\\
         Mean grid power &$\langle{P_\text{grid}}\rangle$& 497 & 505 & 497 & 520 &\si{\kilo\watt}\\
         Total grid energy & $\sum{P_\text{grid}}\Delta t$& 4.37 & 4.43 & 4.37 & 4.57 & \si{\giga\Wh}\\
         Daily average of SA cycles & $\sum|\dot{m}_{sa}|\Delta t/(2M^{\max}_\text{sa} N_\text{days})$& --- & 6.5 & ---& 1.6 & \si{\per\day}\\
         \bottomrule
    \end{tabular}
    \label{tab:table_optimal_energy_power}
\end{table}

\subsection{Net present value}
\begin{figure}
    \centering
    \includegraphics[width=\textwidth]{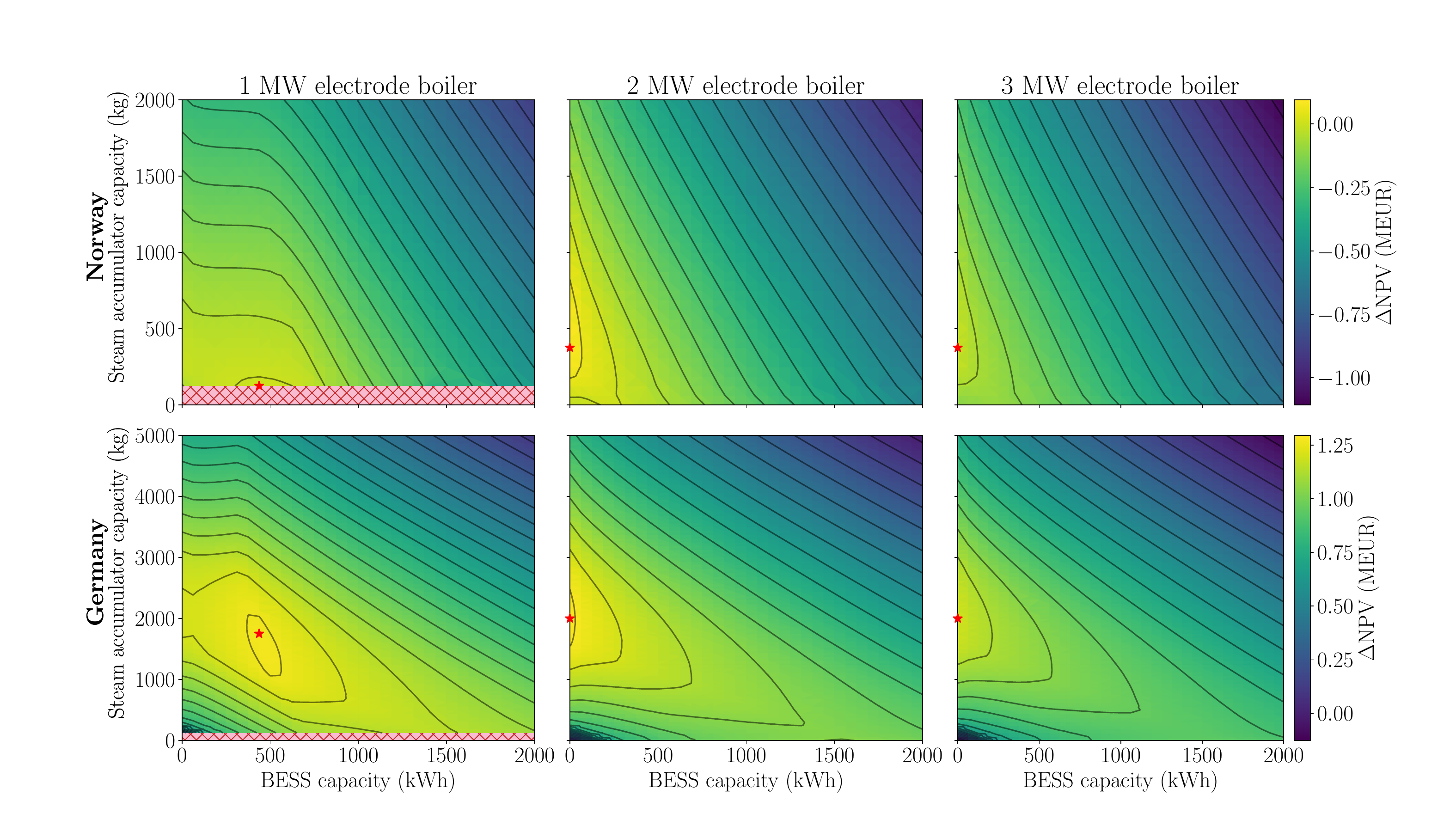}
    \caption{Heat map of the relative net present value ($\Delta$NPV) for Norway (top) and Germany (bottom). NPV as a function of steam accumulator mass capacity and BESS capacity for three different electrode boiler capacities, (a) \SI{1}{\mega\watt}, (b) \SI{2}{\mega\watt}, an (c) \SI{3}{\mega\watt}. The red star indicates the maximum $\Delta$NPV for each capacity of electrode boiler. The red hatched area shows configurations that are not able to meet the steam demand.}
    \label{fig:npv_Peb_Qb_Msa}
\end{figure}

\begin{figure}
    \centering
    \includegraphics[width=\textwidth]{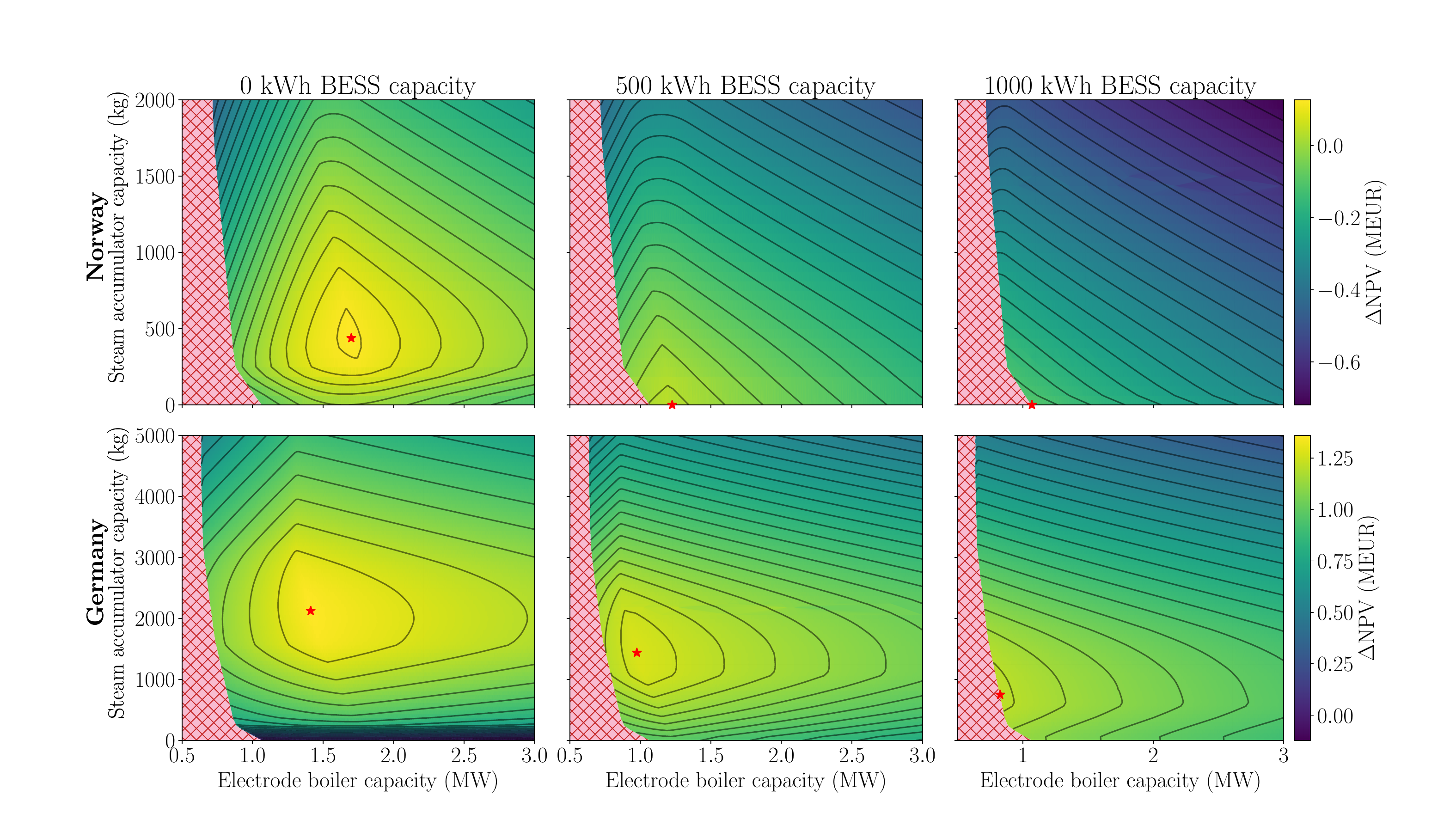}
    \caption{Heat map of the relative net present value ($\Delta$NPV) for Norway (top) and Germany (bottom). $\Delta$NPV as a function of steam accumulator mass capacity and electrode boiler for three different BESS capacities, (a) \SI{0}{\kilo\Wh}, (b) \SI{500}{\kilo\Wh}, an (c) \SI{1000}{\kilo\Wh} and C-rate of 0.9. The red star indicates the maximum $\Delta$NPV for each capacity of electrode boiler. The red hatched area shows configurations that are not able to meet the steam demand.}
    \label{fig:npv_Qb_Peb_Msa}
\end{figure}

\begin{figure}
    \centering
    \includegraphics[width=\textwidth]{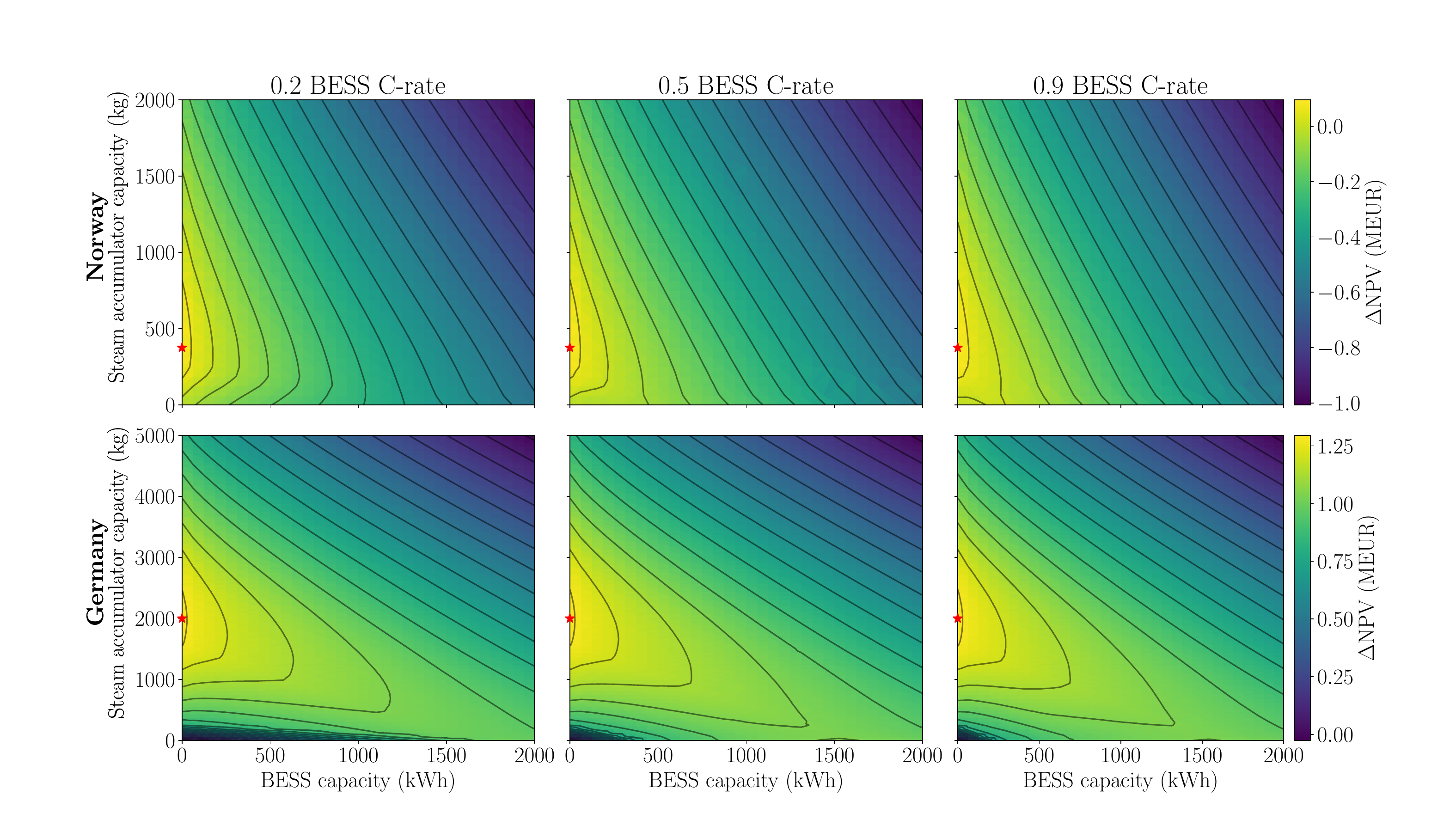}
    \caption{Heat map of the relative net present value ($\Delta$NPV) for Norway (top) and Germany (bottom). NPV as a function of steam accumulator mass capacity and BESS capacity for electrode boiler capacity of \SI{2}{\mega\watt} and three different BESS C-rates (a) 0.2, (b) 0.5, and (c) 1. The red star indicates the maximum $\Delta$NPV for each capacity of electrode boiler.}
    \label{fig:npv_yb_Qb_Msa}
\end{figure}

In figures \ref{fig:npv_Peb_Qb_Msa}, \ref{fig:npv_Qb_Peb_Msa}, and \ref{fig:npv_yb_Qb_Msa}, a heat map of the relative net present value ($\Delta$NPV) is presented as a function of the steam accumulator, electrode boiler, and BESS capacity, and as a function of the BESS C-rate. The relative net present value ($\Delta \text{NPV} = \text{NPV} - \text{NPV}_0$), is calculated as the NPV for the specific configuration minus the NPV for the most cost-efficient configuration with no steam accumulator or BESS installed. Bright colors (yellow) indicate high $\Delta$NPV, while dark (blue) colors indicate small $\Delta$NPV. The color scheme is consistent within the figures for each country, but not between the countries. The red star indicates the local maximum in each graph. The red hatched area shows configurations that are not able to meet the steam demand.

In the figure~\ref{fig:npv_Peb_Qb_Msa}, three electrode boiler sizes are investigated (1, 2, and 3 \si{\mega\watt}) as a function of steam accumulator and BESS capacity. The BESS C-rate is fixed at 0.9. In all cases, a ridge in the $\Delta$NPV is visible with a negative slope. This ridge shows that with increasing BESS capacity, the most cost-efficient solution is to reduce the steam accumulator size. This is expected, as the energy storage on the steam side can be replaced with storage on the electricity side, given that the electrode boiler is large enough.

For all electrode boiler sizes, except for the \SI{1}{\mega\watt} case in Germany, it is clear that having no BESS is the most cost-efficient solution. For \SI{1}{\mega\watt} electrode boiler, it is advantageous to have installed \SI{500}{\kilo\Wh} BESS in Germany. However, this is only a local optimum given at a certain electrode boiler size, and not a global optimal solution. 

Interestingly, most of the configurations for Norway have negative $\Delta$NPV, meaning that the optimal configuration with steam accumulator and electrode boiler has a lower NPV than the configuration with no energy storage installed. This is not observed in Germany, once again showing that the electricity prices in Germany advocate for energy storage, which is not the case for Norway under the applied price assumptions. For an electrode boiler capacity of \SI{1}{\mega\watt}, a red hatched area is visible at low steam accumulator capacities, indicating that these configurations are not able to meet the steam demand. Increasing the BESS capacity for these configurations does not help meeting the steam demand, since the electrode boiler has too low capacity. Steam storage is thus necessary if the boiler cannot meat peak demand

Figure~\ref{fig:npv_Qb_Peb_Msa} shows the $\Delta$NPV for three BESS capacities (0, 500, and 1000~\si{\kilo\Wh}), evaluated as a function of electrode boiler and steam accumulator capacity, while keeping the BESS C-rate fixed at 0.9. Distinct optima are observed in all panels, indicating that a well-defined and cost-effective configuration exists for each BESS capacity. There is clear trend for decreasing steam accumulator size with increasing BESS capacity. This reflects the trade-off between thermal and electrical storage, with BESS partially substituting the need for thermal flexibility when properly sized. However, in several cases, specifically, for 500 and 1000~\si{\kilo\Wh} BESS capacity in Norway and 1000~\si{\kilo\Wh} in Germany, the optimal configuration lies very close to the red hatched area. This region indicates infeasible solutions where the system fails to meet the steam demand. While these border solutions are economically optimal under nominal conditions, they are inherently fragile to uncertainties in steam demand and other unforeseen circumstances. Consequently, configurations farther from the feasibility boundary should be preferred in practice, as they provide valuable operational headroom and increase system robustness. This adds weight to the argument for keeping a certain steam accumulator capacity when combined with small electrode boilers, even when a BESS is present.

Figure~\ref{fig:npv_yb_Qb_Msa} presents the relative net present value ($\Delta$NPV) as a function of steam accumulator and BESS capacity for three different BESS C-rates (0.2, 0.5, and 0.9), with the electrode boiler capacity fixed at \SI{2}{\mega\watt}. The results show that the BESS C-rate has a limited impact on the overall NPV landscape, with only minor shifts in the optimal configuration. However, the slope of the high-value NPV ridge becomes steeper at lower C-rates, most notably in Norway, indicating that lower C-rates require higher BESS capacities to maintain flexibility when steam accumulator capacity is reduced. This effect reflects the reduced power throughput at lower C-rates, which limits the system’s ability to shift energy in response to market signals. Consequently, the choice of C-rate is not a primary driver of system design under current cost assumptions but becomes relevant in scenarios with constrained storage volumes. The profits from FCR is proportional to the C-rate, see supplementary material for detailed results.

\subsection{Utilization of excess heat}

\begin{figure}
    \centering
    \includegraphics[width=0.667\textwidth]{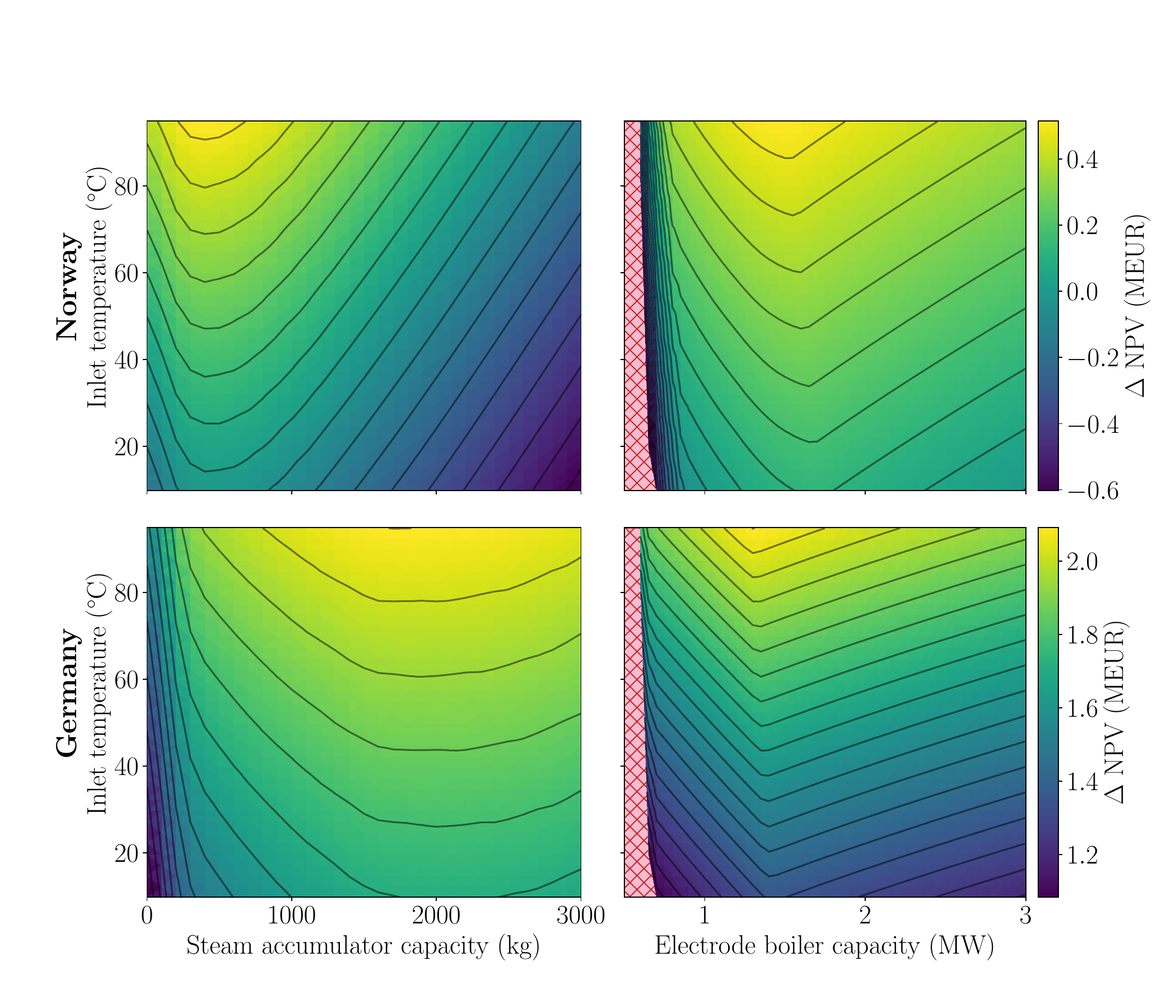}
    \caption{The relative net present value is shown as a function of the inlet temperature to the electrode boiler and steam accumulator capacity with \SI{1.5}{\mega\watt} electrode boiler capacity (left) and electrode boiler capacity with \SI{370}{\kilo\Wh} steam accumulator capacity (right).}
    \label{fig:district_heating}
\end{figure}

The effect of preheating the inlet water to the electrode boiler using excess heat, thereby reducing the energy required to generate steam, is presented in figure~\ref{fig:district_heating}. The origin of the excess heat can be for example from the return loop of the steam network, or it can be waste heat from a nearby factory. The figure presents heat maps of the $\Delta$NPV as a function of inlet water temperature and either steam accumulator capacity (left) or electrode boiler capacity (right) for Norway (top) and Germany (bottom). The $\Delta$NPV is defined relative to a baseline configuration with no energy storage and an inlet temperature of \SI{9.85}{\celsius} (\SI{283}{\kelvin}), which is consistent with the reference used throughout this study. Increasing the inlet temperature to \SI{89.85}{\celsius} results in a substantial improvement in NPV. For Norway, this increase yields an NPV improvement of approximately \SI{0.366}{MEUR} (282\% increase), while in Germany the gain is \SI{0.699}{MEUR} (51\% increase). This gain is primarily driven by the reduced enthalpy requirement to produce steam, which leads to lower electricity consumption and more efficient use of storage. The electrical energy consumption is reduced from \SI{16.0}{\tera\Wh} to \SI{14.4}{\tera\Wh} in Norway (10\% decrease), and from \SI{16.4}{\tera\Wh} to \SI{14.9}{\tera\Wh} in Germany (9.4\% decrease). The maximum grid power is reduced from \SI{851}{\kilo\watt} to \SI{764}{\kilo\watt} (10\% decrease) in Norway, and from \SI{706}{\kilo\watt} to \SI{641}{\kilo\watt} (9.2\% decrease) in Germany. This analysis assumes that sufficient free excess heat is always available. 

The plots also reveal that higher inlet temperatures expand the region of feasible and profitable configurations, shifting the optimal steam accumulator and boiler sizes slightly to the right. In other words, excess heat utilization not only reduces operational costs but also increases the flexibility and operational headroom of the system. In particular, sites with otherwise limited electric boiler capacity or high electricity costs could benefit significantly from thermal preheating.  A more in-depth analysis, including  the additional investment cost and potential excess heat cost, should be done to further investigate the costs and benefits for integrating excess heat utilization for preheating of the inlet water. 

\subsection{Sensitivity of investment cost}

\begin{figure}
    \centering
    \includegraphics[width=\textwidth]{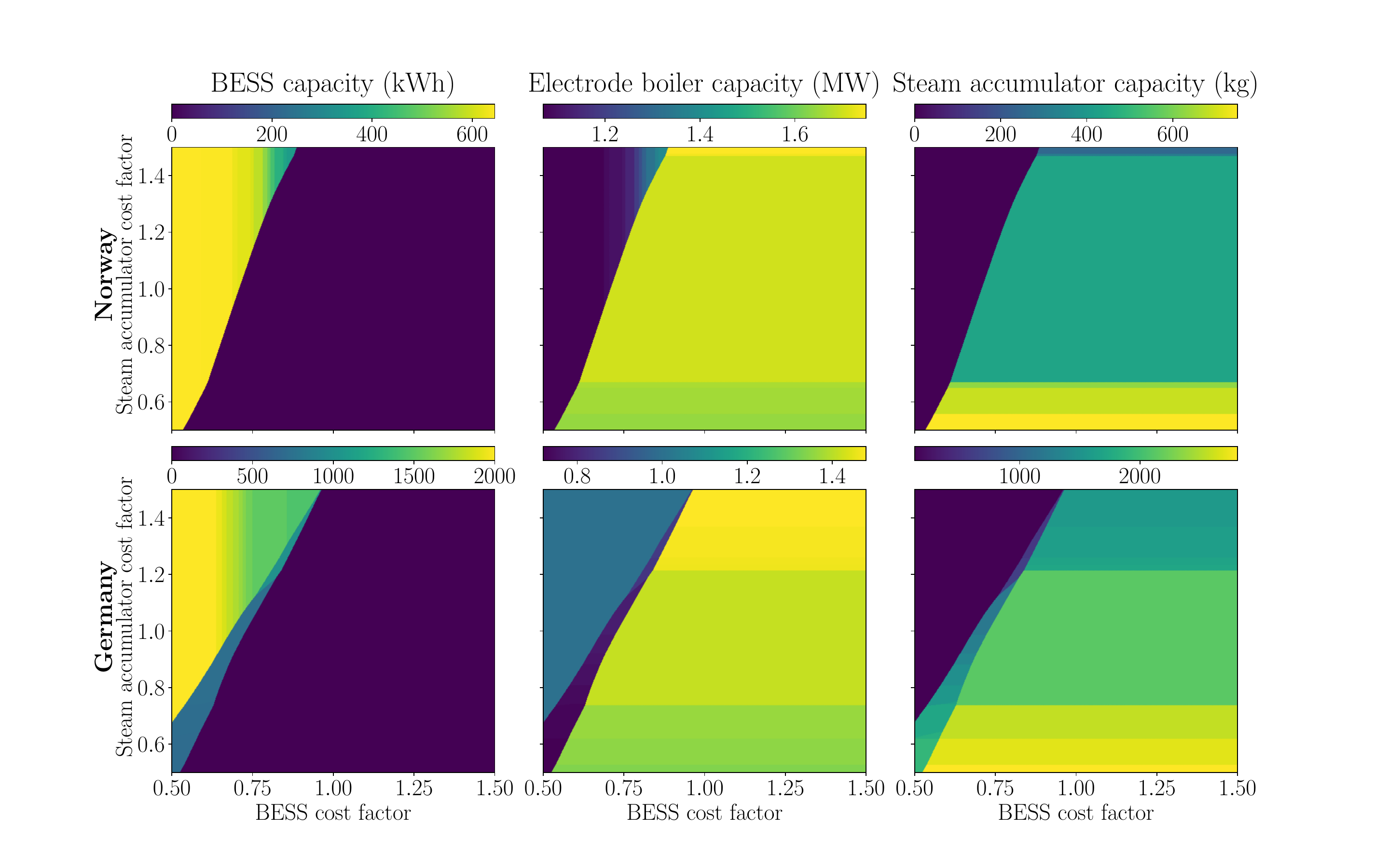}
    \caption{Heat map of the most cost-efficient capacities of (a) BESS, (b) electrode boiler, and (c) steam accumulator for Norway (Top) and Germany (bottom) as a function of BESS and steam accumulator cost factor.}
    \label{fig:cost_sensitivity}
\end{figure}

The sensitivity of the optimal configuration of BESS, electrode boiler, and steam accumulator capacity to changes in investment cost of BESS and steam accumulator is investigated and presented in figure~\ref{fig:cost_sensitivity}. The figure shows heat maps of the most cost-efficient BESS capacity (left), electrode boiler capacity (middle), and steam accumulator capacity (right) for Norway (top) and Germany (bottom), as functions of cost factors for steam accumulator and BESS. The cost factor is defined as $f_i = C'_i/C_i$, where $C'_i$ is the adjusted investment cost and $C_i$ is the baseline cost used throughout this study. For instance, a cost factor of $0.6$ means that the cost is 60\% of the original cost.

The results indicate a clear threshold value: if the BESS cost is reduced below approximately 75\% of its original value ($f_{\text{b}} \approx 0.75$) at the current cost of steam accumulators ($f_{\text{sa}} \approx 1$), a distinct shift occurs where BESS becomes the dominant energy storage solution. Given the strong trend in reducing BESS costs driven by economies of scale and technological progress, these findings suggest that BESS may become increasingly competitive in steam networks in the near future. This underscores the importance of periodically revisiting investment assumptions when designing hybrid thermal-electric energy systems.

\label{sec:sensitivity}

\section{Conclusion}

In this study, we have investigated the economic viability and optimal configuration of a hybrid industrial energy system which can integrate an electrode boiler, steam accumulator, and battery energy storage system (BESS). The system was optimized for minimized net energy cost, using hourly historical data from 2024 for electricity prices, grid tariffs, and frequency containment reserves (FCR) prices in Norway and Germany.

The results show that the economic viability of energy flexibility technologies is highly dependent on regional electricity market structures. In both Norway and Germany, the optimal configuration included a steam accumulator and electrode boiler, but no BESS capacity. This outcome is primarily due to the relatively high investment costs of BESS compared to steam accumulators under current cost assumptions. However, sensitivity analysis on investment costs indicated that a reduction of more than 20\% in BESS cost would make batteries competitive with steam accumulators, particularly in Germany where capacity component of the grid tariffs and electricity price volatility provide stronger incentives for flexible energy use. This result suggest that ongoing cost reductions in BESS technology should be closely monitored when designing future-ready, flexible industrial energy systems.

Significant differences were observed between the two markets. In Germany, high spot prices and high capacity component of the grid tariffs encourage investment in larger steam accumulator capacities, allowing for peak shaving and price arbitrage. In contrast, Norway’s lower capacity component of the grid tariffs result in smaller storage investments with limited economic impact. These findings underscore how market design directly influences the economic attractiveness of different flexibility technologies. For example, increasing the capacity component of the grid tariff will create a strong incentive for technologies that reduce peak power draw.

Revenues from FCR market participation were found to be significant, amounting to approximately 17\% of the net energy costs in Norway and 7\% in Germany under the most cost-efficient configurations. This highlights the importance of integrating ancillary service revenues into economic assessments of industrial flexibility.

In addition it was shown that excess heat integration through electrode boiler inlet water preheating has a substantial value by reducing the overall electricity demand for steam generation and improving system profitability. Although additional investment and operational costs were not included in the analysis, the results point to strong synergies when excess heat is available locally.

Overall, the findings suggest that hybrid steam-electric flexibility systems, especially those centered around steam accumulators, present an economically robust pathway for industrial energy users to reduce costs, enhance operational flexibility, and participate in emerging electricity markets.

\section*{Author contributions}
All authors contributed to formal analysis, investigation, conceptualization, reviewing, and editing. O. G. contributed to methodology, software, writing original drafts, and visualisation. All authors have read and agreed to the published version of the manuscript.

\section*{Acknowledgements}
This research was funded by CETPartnership, the Clean Energy Transition Partnership under the 2024 joint call for research proposals, co-funded by the European Commission (GA N\textsuperscript{o}~101069750) and with the funding organizations detailed on \url{https://cetpartnership.eu/funding-agencies-and-call-modules}. The simulations were performed on resources provided by UNINETT Sigma2, the National Infrastructure for High Performance Computing and Data Storage in Norway.

\section*{Code and Data Availability}
The code used to formulate and solve the optimization model is available at \url{https://gitlab.sintef.no/energy-storage/pflo}. The specific example used in this study is provided in the subfolder \texttt{examples/steam} within the repository. The data used in this study is available at \url{https://doi.org/10.5281/zenodo.15462750}.

\section*{Supplementary material}
Supplementary material is available with results of the grid tariffs, FCR market profits, energy costs, and investment costs as a function of steam accumulator, electrode boiler, and BESS capacity.

\end{document}